\pdfoutput=1

\documentclass[11pt,twoside,a4paper,cmspaper,final,collab]{cms-tdr}
\def\svnVersion{434440P}\def\cmsCernNoTag{CERN-EP-2017-104}\def\cmsCernDate{\today}\def\cmsMessage{Published in Physical Review D as \href{http://dx.doi.org/10.1103/PhysRevD.96.072004}{\doi{10.1103/PhysRevD.96.072004.}}}

\begin{document}\cmsNoteHeader{HIG-15-013}

\hyphenation{had-ron-i-za-tion}
\hyphenation{cal-or-i-me-ter}
\hyphenation{de-vices}
\RCS$Revision: 434329 $
\RCS$HeadURL: svn+ssh://svn.cern.ch/reps/tdr2/papers/HIG-15-013/trunk/HIG-15-013.tex $
\RCS$Id: HIG-15-013.tex 434329 2017-11-15 08:46:40Z sudeshna $
\newlength\cmsFigWidth
\ifthenelse{\boolean{cms@external}}{\setlength\cmsFigWidth{0.85\columnwidth}}{\setlength\cmsFigWidth{0.4\textwidth}}
\ifthenelse{\boolean{cms@external}}{\providecommand{\cmsLeft}{upper}\xspace}{\providecommand{\cmsLeft}{left}\xspace}
\ifthenelse{\boolean{cms@external}}{\providecommand{\cmsRight}{lower}\xspace}{\providecommand{\cmsRight}{right}\xspace}
\ifthenelse{\boolean{cms@external}}{\providecommand{\CL}{C.L.\xspace}}{\providecommand{\CL}{CL\xspace}}
\newcommand{\PLepton}{\ensuremath{\ell}\xspace}
\newcommand{\Pggx}{\ensuremath{\PGg^{*}}\xspace}
\newcommand{\cPX}{\ensuremath{\cmsSymbolFace{X}}\xspace}
\providecommand{\tauh}{\ensuremath{\Pgt_\mathrm{h}}\xspace}
\providecommand{\ptmiss}{\ensuremath{p_{\mathrm{T}}^{\mathrm{miss}}}\xspace}

\cmsNoteHeader{HIG-15-013}
\title{Search for Higgs boson pair production in the \texorpdfstring{$\PQb\PQb\tau\tau$}{b b tau tau} final state in proton-proton collisions at \texorpdfstring{$\sqrt{(s)} =8$\TeV}{sqrt(s) = 8 TeV}}

\date{\today}

\abstract{
Results are presented from a search for production of Higgs boson pairs ($\PH\PH$)
 where one boson decays to a pair of \PQb quarks and the other to a
 $\tau$ lepton pair. This work is based on proton-proton collision data
 collected by the CMS experiment
at $\sqrt{s} = 8\TeV$, corresponding to an integrated
 luminosity of 18.3\fbinv. Resonant and non-resonant
 modes of $\PH\PH$ production have been probed and no significant excess
relative to the background-only hypotheses has been found in either mode.
 Upper limits on cross sections of the two $\PH\PH$ production modes have been set.
The results have been combined with previously
published searches at $\sqrt{s} = 8\TeV$, in decay modes to
 two photons and two \PQb quarks, as well as to four
 \PQb quarks, which also show no evidence for a signal.
 Limits from the combination have been set on resonant $\PH\PH$ production
by an unknown particle \cPX in the mass range
$m_{\cPX}=300\GeV$ to $m_{\cPX}=1000\GeV$.
 For resonant production of spin 0 (spin 2) particles,
the observed 95\% \CL upper limit is
1.13\unit{pb}\,(1.09\unit{pb}) at
$m_{\cPX}=300\GeV$ and to
21\unit{fb}\,(18\unit{fb}) at $m_{\cPX}=1000\GeV$.
 For non-resonant $\PH\PH$ production, a limit of 43 times the rate predicted
 by the standard model has been set.
}

\hypersetup{%
pdfauthor={CMS Collaboration},%
pdftitle={Search for Higgs boson pair production in the bbtautau final state in
 proton-proton collisions at sqrt{(s)} = 8 TeV},%
pdfsubject={CMS},%
pdfkeywords={CMS, physics, Higgs}}

\maketitle

\section{Introduction}
\label{sec:introduction}

The discovery of a standard model (SM)-like Higgs (\PH) boson~\cite{CMS-HIGGS-DISCOVERY,ATLAS-HIGGS-DISCOVERY} motivates further
 investigation of the nature of electroweak symmetry
breaking. In particular,
the measurement of the Higgs self-coupling can provide
 valuable information about the details of the mechanism by which the electroweak symmetry is broken.

\begin{figure}
\centering
\includegraphics[width=0.40\textwidth]{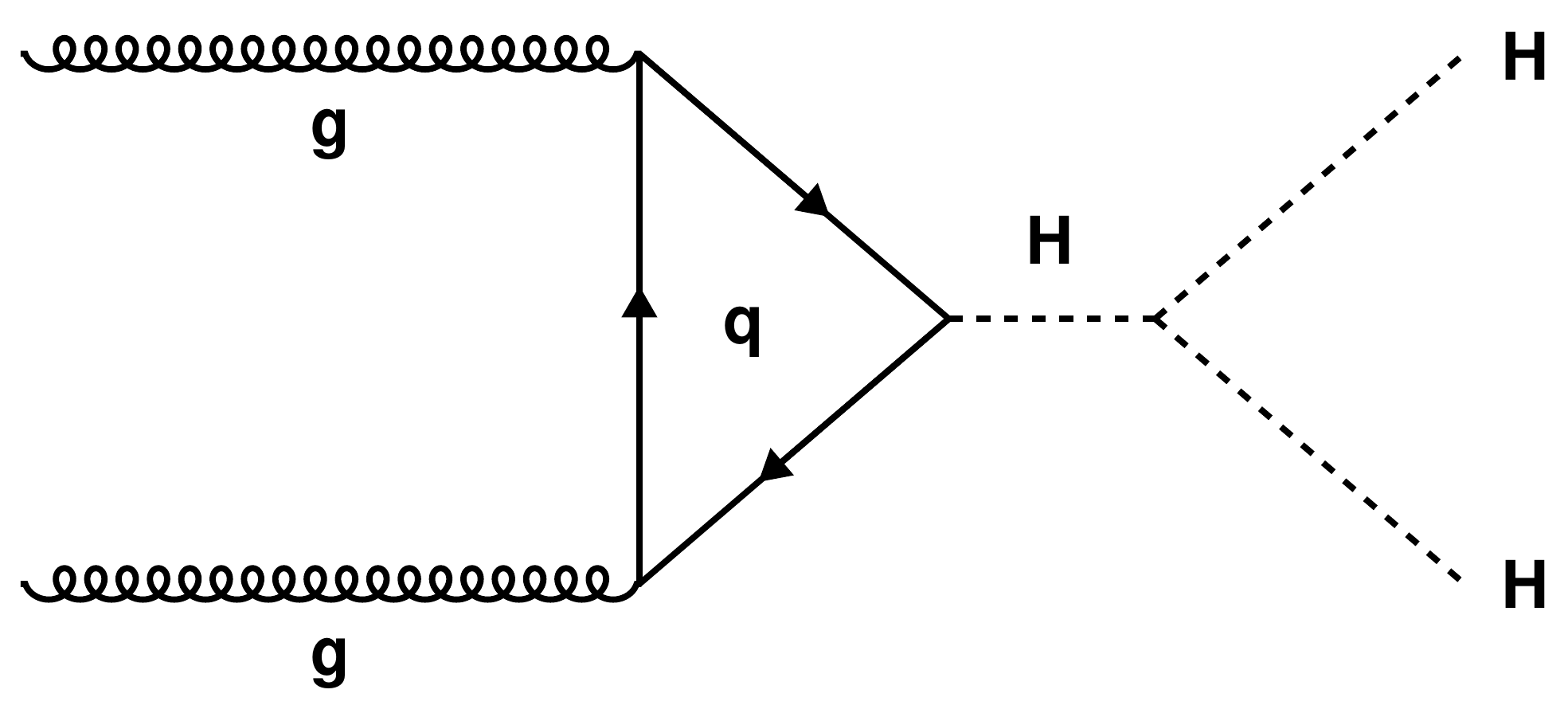} \hspace{\fill}
\includegraphics[width=0.40\textwidth]{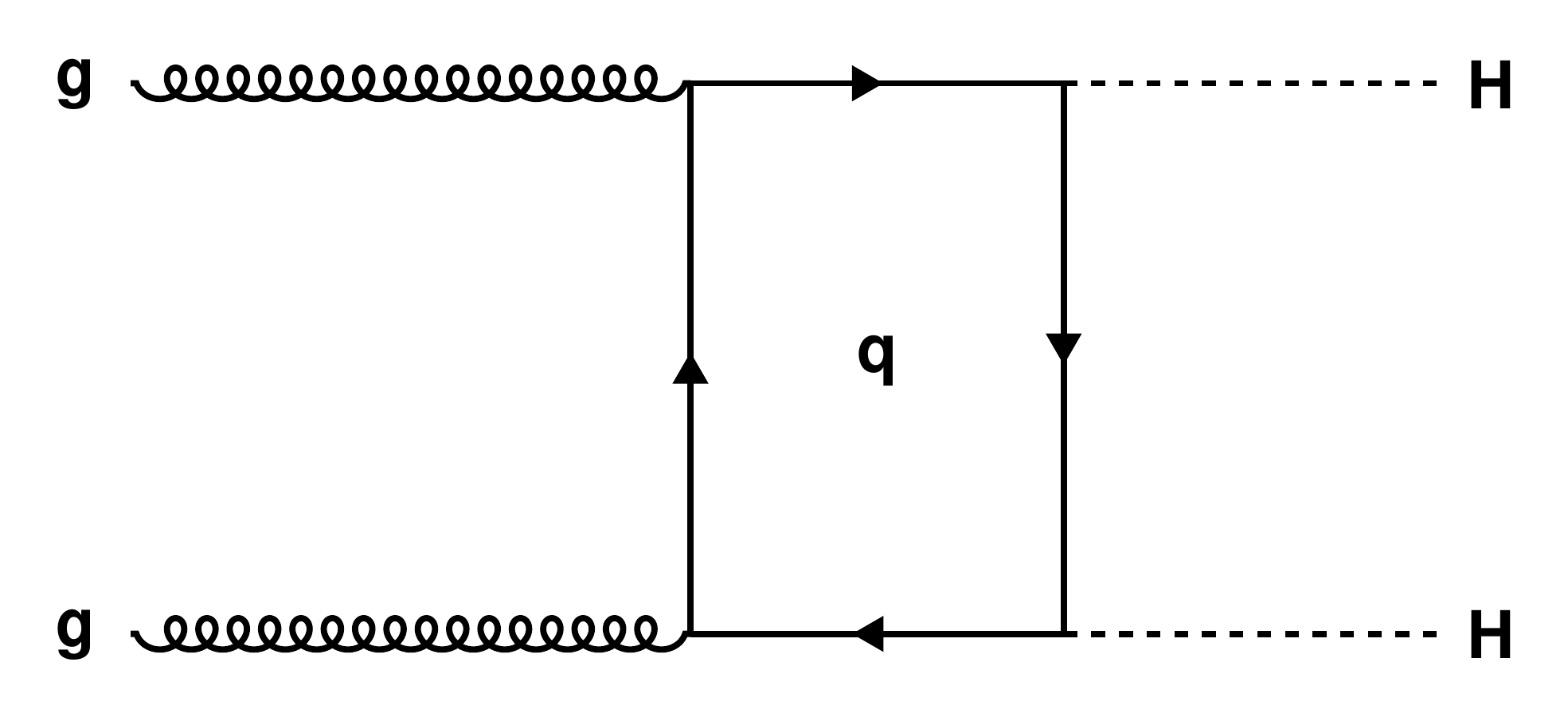} \hspace{\fill}
\caption{ LO Feynman diagrams for $\PH\PH$ production within the SM.}
\label{fig:FeynmanDiagrams_smDiHiggs}
\end{figure}

The measurement of the $\PH$ pair ($\PH\PH$) production rate allows us to probe
 the trilinear $\PH$ self-coupling.
The leading-order (LO) Feynman diagrams for SM $\PH\PH$ production are
 shown in Fig.~\ref{fig:FeynmanDiagrams_smDiHiggs}.
The amplitude of the triangle diagram depends on the trilinear $\PH$ self-coupling. Interference of the box diagram with the triangle diagram
reduces the SM cross
 section to a value of about 10\unit{fb} at a center-of-mass energy of
$\sqrt{s} = 8$\TeV~\cite{deFlorian:2016spz}.
A deviation of the trilinear $\PH$ self-coupling from the SM value may enhance the $\PH\PH$ production rate significantly.
The composite Higgs models discussed in Refs.~\cite{Grober:2010yv, Contino:2012} predict such an enhancement in which the mass distribution of the $\PH$ pair is
 expected to be broad. We refer to this case as non-resonant $\PH\PH$ production.

\begin{figure}
\centering
\includegraphics[width=0.48\textwidth]{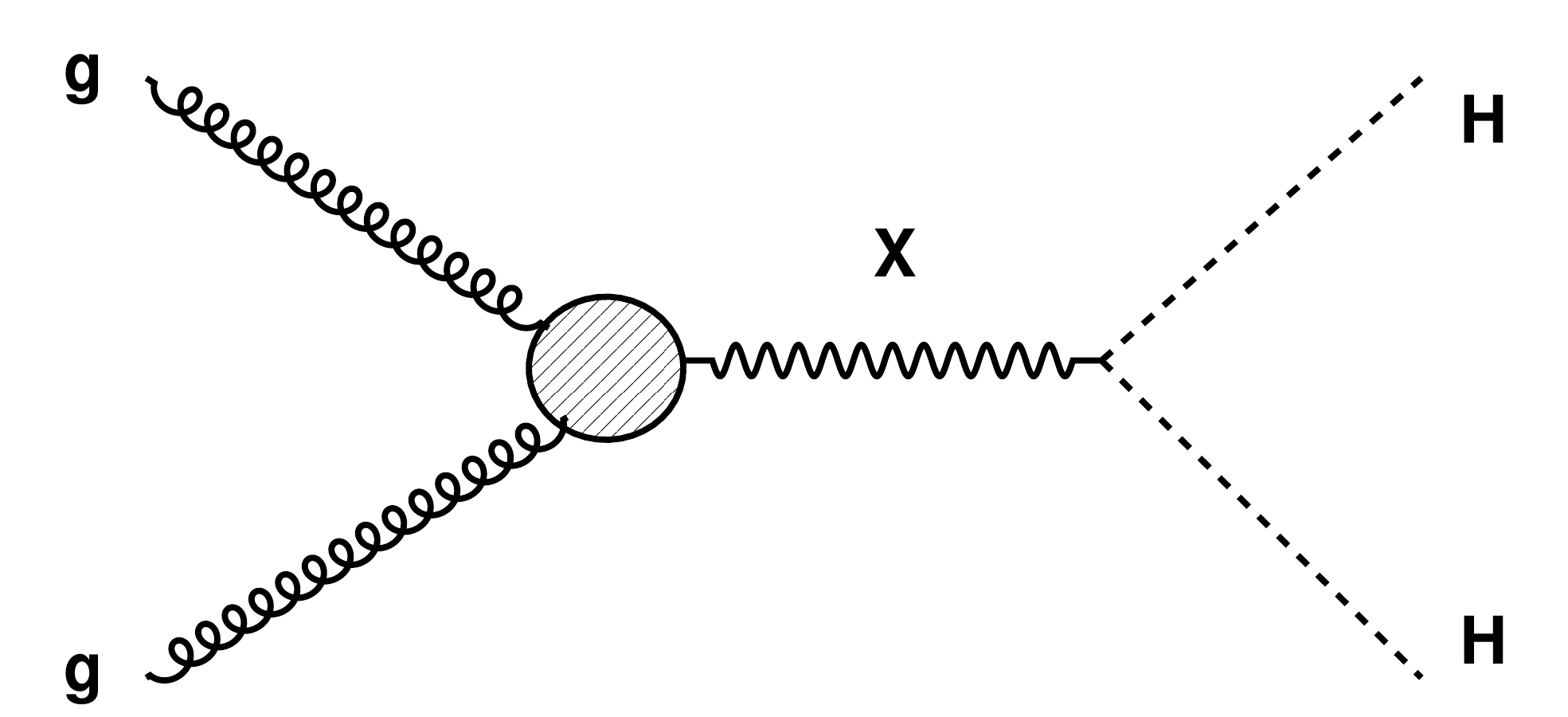}
\caption{
   LO process for the production of a pair of \PH's
 through the decay of a heavy resonance X.}
\label{fig:FeynmanDiagrams_resonant}
\end{figure}

Alternatively, the $\PH\PH$ production rate could be enhanced if a unknown heavy
 particle $\cPX$ decays into a pair of \PH's.
The LO process for this case is shown in
Fig.~\ref{fig:FeynmanDiagrams_resonant}.
We refer to this case as resonant $\PH\PH$ production.
Several models beyond the SM give rise to such decays,
in particular,
two-Higgs-doublet models~\cite{Craig:2013hca,Nhung:2013lpa},
composite Higgs boson models~\cite{Grober:2010yv,Contino:2010mh},
Higgs portal models~\cite{Englert:2011yb,No:2013wsa},
and models involving warped extra dimensions
 (WED)~\cite{Randall:1999ee}.
The present search is performed in the context of the latter models in which the heavy resonance $\cPX$ can either be a radion with spin 0~\cite{Cheung:2000rw,
Gold, DeWolfe, Csaki} or a Kaluza--Klein (KK) excitation of the graviton
with spin 2~\cite{Davoudiasl, Graesser}.
The benchmark points for both models can be expressed in terms of the dimensionless quantity
$k/{{\overline{M}}_{\mathrm{Pl}}}$ and the mass scale
${\Lambda}_{\mathrm{R}} = \sqrt{6}{{\re}^{-kl}}{{\overline{M}}_{\mathrm{Pl}}}$,
where $k$ is the exponential warp factor for the extra dimension, $l$ is the
 size of the extra dimension, and ${\overline{M}}_{\mathrm{Pl}}$ is the
 reduced Planck mass which, is defined by ${M_{\mathrm{Pl}}}/{\sqrt{8{\pi}}}$,
 where ${M}_{\mathrm{Pl}}$ is the Planck mass.
The mass scale ${\Lambda}_{\mathrm{R}}$ is interpreted as the ultraviolet
cutoff of the model~\cite{agashe, aquino}.
In this paper we assume that the SM particles within such a theory follow
the characteristics of the SM gauge group and that the right-handed top quark
 is localized on the \TeV brane, referred to as the elementary top
hypothesis~\cite{BulkRS}. A possible mixing between the
radion and the $\PH$ (r/\PH mixing)~\cite{rHmixing1}
is neglected, since precision electroweak studies show that the
mixing is most likely to be small~\cite{rHmixing2}.

Searches for $\PH\PH$ production have been performed previously by the
CMS Collaboration at the CERN LHC~\cite{CMSsearch_1, CMSsearch_3,
CMSsearch_2b, CMSsearch_2a, CMSsearch_2c} in multi-lepton, multi-lepton$+\gamma\gamma$,
 $\PQb\PQb\tau\tau$, $\gamma\gamma{\PQb}{\PQb}$,
and $\PQb\PQb\PQb\PQb$ final states. In this paper we present
the results for $\PH\PH$ production when one of the \PH's decays to two
bottom quarks, and the other decays to two $\tau$ leptons, where the $\tau$
 leptons decay to hadrons and a ${\nu}_{\tau}$ ($\tauh$). This decay channel
is important because of its large branching fraction. A previous search in
this channel was performed in the mass range of
$m_\cPX=260$--350\GeV~\cite{CMSsearch_3}. The present work extends
 that search to a larger range of resonance mass and to the case of
 non-resonant $\PH\PH$ production. The sensitivity of the analysis is enhanced by
 reconstructing the full four-vector of the $\PH$ that decays into
$\tau$ leptons with a likelihood based algorithm and identifying hadronic
 $\tau$ decays with a multivariate algorithm. We combine the results of the search in the $\PQb\PQb\tau\tau$ decay channel with those from searches
in the $\gamma\gamma{\PQb}{\PQb}$ and
and $\PQb\PQb\PQb\PQb$ final states in order to increase the
sensitivity to potential signals.

The ATLAS Collaboration has searched for
resonant as well as non-resonant $\PH\PH$ production in the $\PQb\PQb\tau\tau$, $\gamma\gamma{\PW}{\PW}^{*}$, $\gamma\gamma{\PQb}{\PQb}$,
 and $\PQb\PQb\PQb\PQb$ decay channels~\cite{ATLASsearch_1, ATLASsearch_2, ATLASsearch_3}.
Their observed (expected) limit on non-resonant $\PH\PH$ production, obtained by combining all channels,
corresponds to 70 (48) times the SM production rate.
The observed (expected) limit on non-resonant $\PH\PH$ production obtained from the $\PQb\PQb\tauh\tauh$ channel alone is 160\,(130) times the rate expected in the SM.
In case of resonant $\PH\PH$ production, the ATLAS Collaboration has set a
combined observed (expected) limit on the production rate
($\sigma(\Pp\Pp \to \cPX) \, \mathcal{B}(\cPX
\to \PH\PH)$)
that ranges from 2.1\unit{pb} (1.1\unit{pb}) at $m_{\cPX} = 300$\GeV to 11\unit{fb}
 (18\unit{fb}) at $m_{\cPX} = 1000$\GeV.
The observed (expected) limit set in the $\PQb\PQb\tauh\tauh$
channel alone ranges from
1.7\unit{pb} (3.1\unit{pb}) at $m_{\cPX} = 300$\GeV to 0.46\unit{pb} (0.28\unit{pb})
 at $m_{\cPX} = 1000$\GeV.

\section{Experimental setup, data, and Monte Carlo events}
\label{sec:datasets_and_MonteCarloSamples}
This Section briefly describes the CMS detector, emphasizing the tracking
 detector which plays an important role in this analysis. Details of the
experimental data set and the Monte Carlo (MC) simulated event samples for
signal events as well as various background processes that are relevant to $\PH\PH$  production and decay are also given here.
\subsection{The CMS detector}
\label{sec:detector}
The central feature of the CMS apparatus is a superconducting
solenoid of 6\unit{m} internal diameter, providing a magnetic field of 3.8\unit{T}. Within
 the superconducting volume are a silicon tracker, a lead tungstate crystal electromagnetic calorimeter,
and a brass and scintillator hadron calorimeter, each composed of a barrel and two endcap sections. In the tracker the inner 3\,(2) layers in the barrel (endcap) region consist of pixel detectors. The outer 10\,(12) layers in the barrel (endcap) region are made of strip detectors. The tracker provides a resolution of $\sim$0.5\% for the measurement of transverse momentum
($\pt$) of tracks which is important for the search described here. Forward calorimeters extend the pseudorapidity ($\eta$) coverage provided by the barrel and endcap detectors. Muons are measured in gas-ionization detectors embedded in the steel flux-return yoke outside the solenoid. A more detailed description of the CMS detector, together with a definition of the
coordinate system used and the relevant kinematic variables, can be found in Ref.~\cite{CMS-JINST}. The CMS trigger system is composed of two levels
~\cite{CMSTrigger}. The first level, composed of custom hardware processors, reduces the event rate from
40\unit{MHz} to 0.1\unit{MHz}. At the next stage, the high-level software-based
trigger, implemented in a farm of about 10\,000 commercial processor cores,
reduces the rate further to less than 1\unit{kHz}.
\subsection{Data and simulated samples}
\label{sec:datasets}
This search is based on proton-proton (pp) collision data corresponding to an integrated luminosity of
 18.3\fbinv recorded at $\sqrt{s} = 8$\TeV in 2012. On average, 21 inelastic pp interactions per LHC bunch crossing occurred during this period~\cite{JME-13-003}. One of the interactions is selected as the primary interaction and the rest are called ``pileup''.
Signal samples for both resonant and non-resonant $\PH\PH$ production are generated with \MADGRAPH 5.1~\cite{MadGraph5}. For resonant $\PH\PH$ production, simulated samples are generated
for spin 0 (radion) and spin 2 (graviton) hypotheses for the $\cPX$ resonance at masses $m_{\cPX} = 300$, 500, 700, and 1000\GeV. Shape templates for the mass parameter of the $\PH\PH$ system used in the signal extraction procedure
 described in Section~\ref{sec:signalExtraction} are produced for intermediate
 mass points using a horizontal template morphing technique~\cite{morphing}
 in steps of 50\GeV between 300 and 700\GeV mass points and in steps of
100\GeV between 700 and 1000\GeV mass points. The efficiency and the acceptance are interpolated linearly between the mass points.

The background contribution from multijet
events is estimated from data, as described in
Section~\ref{sec:backgroundEstimation_QCD}. Background events
arising from $\cPZ/\Pggx \to \PLepton\PLepton$
($\PLepton$ = $\Pe$, $\Pgm$), $\PW$+jets, $\ttbar$, single top quark,
and di-boson ($\PW{\PW}$, $\PW\cPZ$, $\cPZ\cPZ$) production are modeled using
 MC samples. Among these backgrounds $\cPZ/\Pggx \to \PLepton\PLepton$,
$\PW$+jets, $\ttbar$, and di-boson samples
are generated with \MADGRAPH 5.1, while the single top quark samples
are modeled with \POWHEG 1.0~\cite{POWHEG}.

The $\cPZ/\Pggx \to \PLepton{\PLepton}$ and the
 $\PW$+jets backgrounds are generated in bins of generator-level parton
 multiplicity in order to enhance the event statistics in regions of high
 signal purity. These samples are normalized to their respective NNLO
cross sections~\cite{FEWZ}. The $\ttbar$ sample is normalized to the top quark pair production cross section measured by CMS~\cite{CMS-TOP-XSEC-8}
multiplied by a correction factor obtained from a $\ttbar$ enriched control region in data.
Furthermore, a kinematic reweighting is applied to simulated
$\ttbar$ events~\cite{Chatrchyan:2012saa,TOP-12-028} to match the top
quark $\pt$ distribution observed in data.
The single top quark and the di-boson events are normalized to their
respective next-to-leading order (NLO) cross
sections~\cite{MCFMdiBosonXsection}.

Production of events with a single $\PH$ in the SM scenario
 is simulated using \POWHEG 1.0. The production
processes considered are
gluon-gluon fusion ($\Pg\Pg\PH$), vector boson fusion ($\PQq\PQq\PH$), associated production
of the $\PH$ with $\PW$ and $\cPZ$ bosons ($\cmsSymbolFace{V}\PH$),
$\bbbar$ or $\ttbar$ pairs. These samples are produced
for a $\PH$ of mass $m_{\PH} = 125$\GeV
and are normalized to the corresponding cross section given in
Ref.~\cite{LHCHiggsCrossSectionWorkingGroup:2011ti}.
The $\PH$ decays that have been taken into account in this
 analysis are $\PH \to \PQb\PQb$ for $\cmsSymbolFace{V}\PH$ production,
$\PH \to \tau\tau$ for $\cmsSymbolFace{V}\PH$ and $\Pg\Pg\PH$ production,
and both $\PH \to \PQb\PQb$ and $\PH\to \tau\tau$ for
$\PQq\PQq\PH$ production.

Parton shower and hadronization processes are modeled
using \PYTHIA 6.4. Taus are decayed
by \TAUOLA 27.121.5~\cite{tauola}. Pileup interactions represented by minimum bias events generated with
 \PYTHIA 6.4~\cite{pythia6_4} are added to all simulated samples
according to the pileup profile observed in data during the 2012 data-taking
 period. The generated events are
passed through a \GEANTfour~\cite{geant} based simulation of the CMS
detector and are reconstructed using the same version of the CMS software as that for data.

A special technique, referred to as embedding, is used to model the
background arising from $\cPZ/\Pggx \to \Pgt{\Pgt}$ production.
Embedded samples are produced by selecting
$\cPZ/\Pggx \to \Pgm{\Pgm}$ events in data and replacing the reconstructed
 muons by generator-level $\tau$ leptons with the same four-vectors as that
 of the muons~\cite{HIG-13-004}.
The $\tau$ lepton decays are simulated using
\TAUOLA 27.121.5 and their polarization effects are modeled with
 \textsc{TauSpinner} (\textsc{Tauola++} {1.1.4)}~\cite{TauSpinner}.
The visible decay products of the $\Pgt$ are reconstructed with the
particle-flow (PF) algorithm (cf.\ Section~\ref{particleflow}),
and then added to the remaining particles of
the $\cPZ/\Pggx \to \Pgm\Pgm$ event, after removing the two muons. Finally,
the $\tauh$ candidates, the jets, and the missing transverse momentum vector
 $\ptvecmiss$, which is defined as the negative vectorial sum of the 
$\pt$ of all reconstructed particles, are reconstructed, and the event is
analyzed as if it were data. 

The sample of $\cPZ/\Pggx \to \Pgm{\Pgm}$ events that is used as input for the production of $\cPZ/\Pggx \to \Pgt{\Pgt}$ embedded samples
contains contributions from the background $\ttbar \to \PW^{+} \cPqb \, \PW^{-} \cPaqb \to \Pgm^{+}\Pgn_{\Pgm} \cPqb \, \Pgm^{-}\Pagn_{\Pgm} \cPaqb$.
While the overall level of this contribution is small
($\sim$0.1\% of the $\cPZ/\Pggx \to \Pgt\Pgt$ embedded sample),
the contamination of the embedded sample with these events becomes relevant
 for events selected with one or more jets originating from $\PQb$ quarks.
The $\ttbar$ contamination is corrected using simulated
 $\ttbar$ events that are fed through the same embedding procedure as
 described above.

\section{Physics object reconstruction and identification}
\label{src:particleId}
This section describes the methods employed to identify various particles used in this analysis.
\label{particleflow}
The PF algorithm is used to reconstruct and identify individual particles
(referred to as candidates), such as electrons, muons, photons,
charged and neutral hadrons with an optimized combination of
information from various elements of the CMS detector~\cite{CMS-PRF-14-001}.
The resulting candidates are used to reconstruct jets,
hadronic $\tau$ decays, and $\ptvecmiss$. It is required that all
 candidates in an event originate from a common interaction point,
 the primary vertex.
\label{primaryvertex}
The sum of $\pt^2$ of all tracks associated with each interaction
 vertex is computed and the one with the largest value is selected as the
 primary vertex.

\subsection{Jets and \texorpdfstring{$\ptvecmiss$}{missing pt}}
\label{sec:jets}
Jets within $\abs{\eta} < 4.7$
are built using the anti-\kt algorithm~\cite{AntiKT1} implemented in the 
{\FASTJET} package~\cite{AntiKT2},
 with distance parameter of 0.5, using PF candidates as input.
Misreconstructed jets, mainly arising from
 calorimeter noise, are rejected by requiring the jets
to pass a set of loose identification criteria~\cite{jet-fakes}.
Jets originating from pileup interactions are suppressed by an identification
 discriminant~\cite{mvaJetId} based on multivariate (MVA) techniques.
 Corrections based
on the median energy density per event~\cite{Cacciari:2008gn, Cacciari:2007fd}
 as computed by the {\FASTJET} algorithm, are applied to the jet
 energy in order to correct for other pileup effects. The energy of
reconstructed jets is calibrated as a function of $\pt$ and
$\eta$ of the jet~\cite{Chatrchyan:2011ds}.
Jets of $\abs{\eta} < 2.4$ and $\pt > 20$\GeV are tagged as
$\PQb$ quark jets if they are selected by an MVA based algorithm
which uses lifetime information of $\PQb$ quarks
(``Combined Secondary Vertex'', CSV, algorithm). The $\PQb$ tagging
 efficiency and mistag (misidentification of jets without $\PQb$ quarks
as $\PQb$ quark jets) rates for this search are 70\% and 1.5\% (10\%)
for light (charm) quarks respectively~\cite{Chatrchyan:2012jua}.

\label{sec:MET}

The magnitude and direction of the $\ptvecmiss$ vector are reconstructed
 using an MVA based algorithm~\cite{JME-13-003} which uses the fact that pileup predominantly produces low-$\pt$ jets and `unclustered energy' (hadrons not within jets),
while isolated leptons and high-$\pt$ jets are almost exclusively produced by the hard-scatter interaction, even in high-pileup conditions. In addition, the algorithm provides event-by-event estimate of the $\ptvecmiss$ resolution.

\subsection{Lepton identification}
\label{sec:electronAndMuonIdAndIso}
Electrons and muons are used in this analysis solely for the purpose of
vetoing events, as described in Section~\ref{sec:eventSelection}. A
description of the electron and the muon identification criteria and the
computation of their isolation from other particles is given in
Refs.~\cite{Khachatryan:2015hwa, Chatrchyan:2012xi}.
\label{sec:tauId}

 The reconstruction
 of a $\tauh$ lepton starts with a PF jet as the initial seed. This is
 followed by
the reconstruction of the ${\Pgp}^0$ components in the jet which are then
combined with the charged hadron components
 to fully reconstruct the decay mode of the $\tauh$ and to calculate its
four-momentum~\cite{TAU-11-001}.
The identification of $\tauh$ is performed by a MVA based
discriminant~\cite{tauid}.
The main handle to separate hadronic $\tau$ decays
from quark and gluon jets is the isolation of the $\tauh$ candidate
from other charged hadrons and photons.
Variables that are sensitive to the distance of separation between the
production and decay vertices of the $\tauh$ candidate
 complement the MVA inputs.
This algorithm achieves a $\tauh$ identification efficiency of 50\%
 with a misidentification rate for quark and gluon jets below 1\%.
Additional discriminants are used to separate
$\tauh$ candidates from electrons and muons~\cite{tauid}. The discriminant
 against electrons
uses variables sensitive to electron shower shape, electron track,
and $\tauh$ decay kinematics. The discriminant against muons uses inputs based
 on calorimetric information of the $\tauh$ jet and reconstructed hits and
 track segments in the muon system.

\section{\texorpdfstring{$\PH\PH$}{HH} mass reconstruction and event selection}
\label{sec:eventSelection}

\label{sec:massReconstruction}

This analysis is based on data satisfying a $\tauh\tauh$ trigger which
 requires  the presence of two $\tauh$ objects with a
$\pt$ threshold of 35\GeV and $\eta \leq 2.1$ for each $\tauh$.
A further selection of events is made offline. It is first ensured
that the data considered in the analysis are of good quality and each event
 contains a primary vertex with the absolute value of the $z$ coordinate less
 than 24\unit{cm}, and within the radial distance of 2\unit{cm} from the beam axis.
The following
analysis specific selection criteria are then applied, determined by the need
 to suppress specific types of backgrounds. These selection criteria
 depend on the mass of the pair of $\tauh$ candidates and the pair of
\PQb quark jets which are determined as follows.

The $\PH$ that decays into a pair of $\tauh$ leptons is
 reconstructed by a likelihood based algorithm, referred to as {SVfit}~\cite{Bianchini:2014vza}. The
algorithm uses the four-momenta of the two $\tauh$ candidates, the
magnitude and direction of the $\ptvecmiss$ vector as well as the event-by-event estimate of the $\ptvecmiss$ resolution
as input to reconstruct the full
four-momentum vector ($\pt$, $\eta$, $\phi$, and mass) of the pair of $\tauh$
candidates without any constraint on its mass.
A mass window constraint is later applied as described below.
The four-vector of the $\PH$ that decays into $\PQb$ quarks is
 reconstructed by means of a kinematic fit. The fit varies the energy of
the highest quality (according to the CSV algorithm)
$\PQb$ quark jet within the expected
resolution, keeping the jet direction fixed, subject to the constraint that
 the invariant mass of the two $\PQb$ quark jets equals
$m_{\PH} = 125$\GeV. Further selection is based on a mass window
criterion as described below.

In the search for resonant $\PH\PH$ production, the four-momentum vectors of the two \PH's are used to reconstruct the mass of the $\PH\PH$ system,
$m_{\PH\PH}$.
We assume that the width of the new particle $\cPX$ is small compared to the experimental resolution on the mass of the $\PH$ pair,
which, for resonances of true mass $m_{\cPX}$ in the range 300\GeV to 1000\GeV,
typically amounts to 8\% times $m_{\cPX}$.
A peak in the $\PH\PH$ mass distribution is expected this case.
The search for heavy spin 0 and spin 2 resonances is hence based on finding a peak in the $\PH\PH$ mass spectrum.

In the non-resonant case, the mass distribution of the $\PH$ pair is expected to be broader than the experimental resolution.
 After comparing different observables in terms of their capability to
 separate a potential signal from the background we have found that the
observable $m_{\mathrm{T2}}$~\cite{Lester:1999tx} performs the best.
Our search for non-resonant $\PH\PH$ production is hence based on the
$m_{\mathrm{T2}}$ variable
 which is an analog of the transverse mass variable used
in $\PW\to \ell \nu$ analyses, adapted to the cascade decays
of $\ttbar$ pairs to pairs of \PQb quarks, leptons, and neutrinos. It
improves the separation of the $\PH\PH$ signal in particular from the $\ttbar$ background, due to the fact that
values of the $m_{\mathrm{T2}}$ variable extend up to 300--400\GeV for
signal events,
while for $\ttbar$ background events they are concentrated below the top
quark mass. The usage of this observable in analyses of non-resonant $\PH\PH$  production in the $\PQb\PQb\Pgt\Pgt$ final state was first proposed
 in Ref.~\cite{mT2}.

The selection of events is based on the following additional requirements:
\begin{itemize}
\item The event is required to contain two $\tauh$ candidates with
$\pt > 45$\GeV and $\abs{\eta} < 2.1$, which pass the
identification criteria described in Section~\ref{sec:tauId}.
 Both $\tauh$ candidates are required to be
matched to the $\tau$ objects that trigger the event within
$\Delta{R} < 0.5$. Here
$\Delta{R}$ = $\sqrt{\smash[b]{(\Delta{\eta})^2 + (\Delta{\phi})^2}}$ and $\Delta{\eta}$
 and $\Delta{\phi}$ are the distances in pseudorapidity and azimuthal
 angle (in radians), respectively, between the reconstructed tau object and
the tau object at the trigger level.
\item The two $\tauh$ candidates are required to be of opposite charge.
  The $\tauh\tauh$ invariant mass ($m_{\tau\tau}$),
 reconstructed by the {SVfit} algorithm,
  is required to be in the window 80--140\GeV. If
  multiple combinations exist in an event,
  the combination with the highest sum of outputs from
the MVA based discriminant
  that separates the $\tauh$ candidate from quark and gluon jets, is taken.
\item The event is required to contain two jets of $\pt > 20$\GeV and
$\abs{\eta} < 2.4$.
  The jets are required to be separated from each of the two $\tauh$
 candidates by $\Delta R > 0.5$. The mass of the two jets
  is required to be within the window $80 < m_{\textrm{jj}} < 170$\GeV.
\item Events containing an isolated electron of $\pt > 15$\GeV and
 $\abs{\eta} < 2.4$, or an isolated muon of $\pt > 15$\GeV
and $\abs{\eta} < 2.4$ are rejected.
\end{itemize}

In the search for non-resonant $\PH\PH$ production, the Lorentz boost of the \PH's
 and the resulting boost of the $\tauh$ lepton pair coming from their
decays is used to further distinguish between signal and background events by
requiring the distance in $\eta$-$\phi$ between the two $\tauh$ candidates,
$\Delta R_{\tau\tau}$, to be less than 2.0. This criterion is not used in
the resonant $\PH\PH$ search in order to preserve sensitivity in the low mass
($m_{\PH\PH}<500$\GeV) region. Except for the $\Delta R_{\tau\tau}$ criterion,
 the event and object selection applied in the search for non-resonant and
 for resonant HH production are identical.

\section{Definition of event categories}
\label{sec:eventCategories}

The $\PH\PH \to \PQb\PQb\Pgt\Pgt$ signal events are expected
to contain two $\PQb$ quark jets in the final state. The efficiency to
reconstruct a single \PQb jet is higher than reconstructing two \PQb jets in an
 event. The efficiency of signal selection is therefore enhanced in this
analysis by accepting events with one $\PQb$ tagged jet and one jet
 which is not $\PQb$ tagged. A control region containing events with
 two or more jets, none of which passes the $\PQb$ tagging criteria,
 is used to constrain systematic uncertainties. More specifically, the event categories
 are:
\begin{itemize}
\item {2 \PQb tags} \\
  Events in this category are required to contain at least two jets of $\pt > 20$\GeV and $\abs{\eta} < 2.4$ which are selected by the CSV discriminant described in Section~\ref{sec:jets}.
\item {1 \PQb tag} \\
  Events in this category are required to contain one jet of $\pt > 20$\GeV and $\abs{\eta} < 2.4$, which is selected by the CSV discriminant and
 one or more additional jets of $\pt > 20$\GeV. These jets are required
to either not satisfy $\abs{\eta} < 2.4$ or not to be selected by the
CSV discriminant.
\item {0 \PQb tags} \\
  Events in this category are required to contain at least two jets of
$\pt > 20$\GeV, all of which either do not satisfy
$\abs{\eta} < 2.4$ or are not selected by the CSV discriminant.
\end{itemize}
These categories are mutually exclusive. For the purpose of studying the modeling of data by MC simulation
in a region that is not sensitive to the presence or the absence of signal
events, we define as `inclusive' category the union of all three
categories. No selection criteria are applied on
$m_{\Pgt\Pgt}$, $m_{\cmsSymbolFace{jj}}$, or
$\Delta R_{\Pgt\Pgt}$ in the inclusive category.

\section{Background estimation}
\label{sec:backgroundEstimation}
The two important sources of background in the 0 \PQb tag and 1 \PQb tag
 categories are events containing $\cPZ/\Pggx \to \Pgt\Pgt$ decays
 and multijet production. In the 2 \PQb tag category
$\cPZ/\Pggx \to \Pgt\Pgt$ decays and $\ttbar$ events are dominant
sources of background events.

\subsection{The multijet events}
\label{sec:backgroundEstimation_QCD}

The reconstructed $\tauh$ candidates in multijet events are typically
due to the misidentification of quark or gluon jets.
The contribution from this background in the signal
region, in terms of event yield and shape of the distributions in $m_{\PH\PH}$ and $m_{\mathrm{T2}}$ (``shape template''),
is determined entirely from data.
The normalization and shape is obtained separately in each event category,
from events that pass the selection criteria described in
Sections~\ref{sec:eventSelection} and contain two $\tauh$ candidates of
opposite charge. It is required that the leading (higher $\pt$) $\tauh$ candidate passes relaxed,
but fails the nominal $\tauh$ identification criteria.
The probabilities for the leading $\tauh$ candidate to pass the relaxed and nominal $\tauh$ identification criteria
are measured in events that contain two $\tauh$ candidates of the same charge,
as functions of
 $\pt$ of the leading $\tauh$ candidate in three regions of
 $\eta$, $\abs{\eta} < 1.2$, $1.2 < \abs{\eta} < 1.7$, and
 $1.7 < \abs{\eta} < 2.1$. A linear function is fitted to the variation
of the ratio of these two probabilities with $\pt$ and
 is applied as an event weight to obtain the estimate for
 the shape template of the multijet background in the signal region. Contributions from other backgrounds to these
events are subtracted based on MC predictions.

\subsection{The \texorpdfstring{$\cPZ/\Pggx \to \Pgt\Pgt$}{Z to gamma* tau tau} events}
\label{sec:backgroundEstimation_Ztautau}

The dominant irreducible $\cPZ/\Pggx \to \Pgt\Pgt$ background in the
 event categories with 2 \PQb tags, 1 \PQb tag, and 0 \PQb tags is modeled
by applying embedding to $\cPZ/\Pggx \to \Pgm\Pgm$ events selected from data
as described in Section~\ref{sec:datasets}. The embedded sample is normalized
to the $\cPZ/\Pggx \to \Pgt\Pgt$ event yield obtained from the MC simulation
in the inclusive event category.
The correction due to  $\ttbar$ contamination is performed by
subtracting the distribution in $m_{\PH\PH}$ or $m_{\mathrm{T2}}$
whose shape and normalization are determined using the $\ttbar$
embedded sample from that in the $\cPZ/\Pggx \to \Pgt\Pgt$ embedded sample
 in each event category. An uncertainty on the number of events in each bin
 is set to the sum of
uncertainties of the $\cPZ/\Pggx \to \Pgt\Pgt$ and
 $\ttbar$ embedded yields in that bin, added in quadrature.

  The embedded samples cover only a part of the $\cPZ/\Pggx \to \Pgt\Pgt$
background, namely events in which both reconstructed $\tauh$ candidates match generator-level hadronic $\tau$ decays, because of
 requirements that are applied at the generator level during the production of the embedded samples to enhance the number of events that pass the selection criteria described in Sections~\ref{sec:eventSelection}. The small
additional contribution arising from $\cPZ/\Pggx \to \Pgt\Pgt$ production in which one or both reconstructed $\tauh$ candidates are due to a
 misidentified electron, muon, or jet are taken from the
$\cPZ/\Pggx \to \Pgt\Pgt$ MC sample.

\subsection{Other backgrounds}
\label{sec:backgroundEstimation_others}
The contribution of $\ttbar$ background is estimated using an MC sample
 after reweighting the events as described in Section~\ref{sec:datasets}.
The background contributions arising from $\PW$+jets,
 $\cPZ/\Pggx \to \PLepton\PLepton$ ($\PLepton= \Pe$, $\Pgm$),
single top quark, and di-boson
production, as well as from the production of events with a single SM H boson
 are small and are modeled using MC samples.

\section{Systematic uncertainties}
\label{sec:systematicUncertainties}
The systematic uncertainties in this analysis may affect the number of signal
 or background events selected in a given event category or affect the relative number of signal or background
 events in individual bins of kinematic distributions.
An additional uncertainty arises due to the limited statistics
available to model the $m_{\PH\PH}$ or $m_{\mathrm{T2}}$ distributions of individual backgrounds in some of the event categories. The treatment of such uncertainties is described in Section~\ref{sec:signalExtraction}.
\label{sec:systematicUncertainties_yield}
The systematic uncertainties relevant to this analysis are
\begin{itemize}
\item {The $\tauh$ trigger and identification efficiency} \\
  The uncertainty in the $\tauh$ identification efficiency has
  been measured as 6\% using $\cPZ/\Pggx \to \Pgt\Pgt \to \mu\tauh$
 events. The $\tauh$ candidates in $\cPZ/\Pggx \to \Pgt\Pgt$ events typically
  have $\pt$ in the range 20 to 50\GeV. An uncorrelated uncertainty of
 $20\%  \pt/(1000\GeV)$
is added to account for the extrapolation to the high-$\pt$ region,
including the uncertainty in the charge misidentification rate of
high-$\pt$ $\tau$ leptons. The above uncertainties have been taken from
Ref.~\cite{tauid}.
  The uncertainty in the efficiency of the $\tauh\tauh$ trigger
  amounts to 4.5\% per $\tauh$ candidate~\cite{CMSsearch_3}.
\item {\tauh energy scale} \\
 The uncertainty in the $\tauh$ energy scale is taken as 3\%~\cite{tauid}.
\item {Background yields} \\
  The rate of the $\cPZ/\Pggx \to \PLepton\PLepton$ ($\PLepton = \Pe$, $\Pgm$) background is attributed an uncertainty of 5\%.
  The normalization of the $\cPZ/\Pggx \to \Pgt\Pgt$ embedded samples, as described in Section~\ref{sec:backgroundEstimation_Ztautau}, is attributed an uncertainty of 5\%.
  An additional uncertainty of 5\% is assigned to the fraction of
$\cPZ/\Pggx \to \Pgt\Pgt$ events
  entering the 2 \PQb tags and 1 \PQb tag categories. This uncertainty has been
introduced to cover potential small biases of the embedding technique. The rate of the $\cPqt\cPaqt$ background is known with an uncertainty of 7\%.
  The uncertainty in the MC yield of single top quark and di-boson backgrounds amounts to 15\%. An uncertainty of  30\% has been applied to the $\PW$+jets background yield obtained from MC. The above uncertainties
have been taken from Refs.~\cite{HIG-13-021, CMSsearch_3}.
\item {Integrated luminosity} \\
  The uncertainty in the integrated luminosity is taken as 2.6\%~\cite{Lum}.
  This uncertainty is applied to signal
  and to $\cPZ/\Pggx \to \PLepton\PLepton$ ($\PLepton = \Pe$, $\Pgm$,
$\Pgt$), $\PW$+jets, single top quark and di-boson backgrounds.
This uncertainty is not applied to the
 $\ttbar$ background, as this background is normalized to the top quark
 pair production cross section measured by CMS with a correction factor obtained from
 a $\ttbar$ dominated control region
 in data as described in Section~\ref{sec:datasets}. The normalization of the multijet background is obtained from data and hence is not subject to the luminosity uncertainty.

\label{sec:systematicUncertainties_shape}

\item {Jet energy scale} \\
  Jet energy scale uncertainties range from 1 to 10\% and are parametrized as
functions of jet $\pt$ and $\eta$ ~\cite{Chatrchyan:2011ds}.
  They affect the yield of signal and background events in different event categories and the shape of the $m_{\PH\PH}$ and $m_{\mathrm{T2}}$
distributions.
\item {The \PQb tagging efficiency and the mistag rate} \\
  Uncertainties in the $\PQb$ tagging efficiencies and the mistag rates
result in event migration between categories. These are evaluated as
 functions of jet $\pt$ and $\eta$ as determined in
Ref.~\cite{Chatrchyan:2012jua} and are applied to MC samples.
\item {The multijet background estimation} \\
  The uncertainty in this background contribution is obtained by adding the
  statistical uncertainty in the yield of events in the sample with two
opposite charge $\tauh$ candidates in quadrature with the uncertainty in
the slope and offset parameters of the function used as event weight to the
shape template as described in
Section~\ref{sec:backgroundEstimation_QCD}.
\item {The $\ptvecmiss$ resolution and response} \\
 The uncertainties related to the magnitude and direction of the $\ptvecmiss$ vector, which affect the shape of the
 $m_{\PH\PH}$ and $m_{\mathrm{T2}}$ distributions, are covered by
 uncertainties in the
$\cPZ$ boson recoil correction. The $\cPZ$ boson recoil correction is
 computed by comparing data with simulation in $\Z\to
\Pe\Pe$, $\Z\to \PGm\PGm$, and photon+jets samples,
which do not have any genuine missing transverse momentum.
All observables related to $\ptvecmiss$
(including $m_{\PH\PH}$ and $m_{\mathrm{T2}}$) are recomputed by
varying $\ptvecmiss$ within its uncertainty~\cite{JME-13-003} and applied
 to MC samples.
\item {The top quark $\pt$ reweighting} \\
  The reweighting that is applied to simulated $\ttbar$
  events (Section~\ref{sec:datasets}) is varied between
  one (no correction) and twice the reweighting factor
 (overcorrection by 100\%) to account for the uncertainty due to reweighting~\cite{TOP-12-028, Chatrchyan:2012saa}.
\item {Other sources}\\
The uncertainties on the SM $\PH\PH$ cross section are $+4.1$\%/$-5.7$\% due to scale,
 $\pm$5\% due to approximations concerning top quark mass effects that are made in the theoretical calculations, $\pm$2.6\% due to ${\alpha}_S$ and $\pm$3.1\% due to
the parton density function~\cite{deFlorian:2016spz}. The uncertainty due to the
 $\PH\to{\tau\tau}$ ($\PH\to{\PQb\PQb}$)
branching fraction is $\pm$3.3\% ($\pm$3.2\%)~\cite{Heinemeyer:2013tqa}.
The effect of the uncertainty on the number of pileup interactions amounts to less than 1\% and is neglected.
\end{itemize}

\section{Signal extraction}
\label{sec:signalExtraction}

Signal rates are determined from a binned maximum
likelihood fit for signal plus background and background-only hypotheses.
In case of resonant (non-resonant) $\PH\PH$ production, we fit the distribution
of $m_{\PH\PH}$ ($m_{\mathrm{T2}}$),
reconstructed as described in Section~\ref{sec:massReconstruction}.
Constraints on systematic
uncertainties that correspond to multiplicative factors on the signal or
the background yield
(e.g. cross sections, efficiencies, misreconstruction rates, and
sideband extrapolation factors) are represented by log-normal
probability density functions.
Systematic uncertainties in the shape of $m_{\PH\PH}$ and $m_{\mathrm{T2}}$
distributions for signal as well as background processes are accounted for
by the `vertical template morphing' technique ~\cite{Conway} and represented by Gaussian probability density functions. The
Barlow--Beeston method~\cite{BarlowBeeston, Conway} is employed to account
 for statistical uncertainties on the $m_{\PH\PH}$ and
$m_{\textrm{T2}}$ shape templates.

\section{Results}
\label{sec:results}

\subsection{Observed yields}
\label{sec:resultsYields}

The number of events observed in the event categories with 2 \PQb tags, 1 \PQb tag, and 0 \PQb tags as
 well as the expected yield of background processes in these categories are
given in Table~\ref{tab:eventYield}. The signal rate expected for
non-resonant $\PH\PH$ production has been computed for a cross section
$\sigma(\Pp\Pp \to \PH\PH)$ of 1\unit{pb},
 corresponding to 100 times the SM cross section, and SM event kinematics~\cite{Oliveira, Barger}.  In the case of resonant $\PH\PH$ production,
the signal yield has been computed for a resonance $\cPX$ (radion or graviton)
 of mass $m_{\cPX} = 500$\GeV and a
$\sigma(\Pp\Pp \to \cPX) \, \mathcal{B}(\cPX
\to \PH\PH)$ of 1\unit{pb}. The corresponding WED
 model parameters are
$kl= 35$, $k/{{\overline{M}}_{\mathrm{Pl}}}$ = 0.2, assuming an elementary
top hypothesis and no radion--Higgs (r/\PH)
mixing~\cite{BulkRS, rHmixing1, rHmixing2}.

\begin{table*}[htbp]
\newcolumntype{x}{D{,}{\,\pm\,}{4.3}}
\topcaption{
  Observed and expected event yields in different event categories, in the
 search for non-resonant (top) and resonant (bottom) $\PH\PH$ production
(($\Pp\Pp \to \cPX) \, \mathcal{B}(\cPX \to \PH\PH)$).
Expected event yields are computed using
 values of nuisance parameters obtained by the maximum likelihood fit to the
data as described in Section~\ref{sec:signalExtraction}. Quoted uncertainties
 represent the combination of statistical and systematic uncertainties.
The WED model parameters are $kl = 35$, $k/{{\overline{M}}_{\mathrm{Pl}}}$
= 0.2 (assuming an elementary top hypothesis and no radion--Higgs mixing).
}
\centering
\begin{scotch}{lxxx}
\multicolumn{4}{c}{Non-resonant analysis (event yields)} \\
\hline
Process & \multicolumn{1}{c}{0 \PQb tags} & \multicolumn{1}{c}{1 \PQb tag} & \multicolumn{1}{c}{2 \PQb tags} \\
\hline
Non-resonant $\PH\PH$ production (100 SM)   &                 1.2,0.2 & 4.6,0.6 & 4.3,0.5\\
\hline
$\cPZ \to \Pgt\Pgt$                      &       120.3,11.1 & 17.7,3.0 & 2.0,0.8\\
Multijet                                     & 27.9,2.7  & 5.4,1.0 & 0.7, 0.2 \\
$\PW$+jets                                       & 4.3, 0.8  & 0.4,0.1 & 0.4,0.1 \\
$\cPZ$+jets ($\Pe$, $\Pgm$, or jet misidentified as $\tauh$) & 0.7,0.2  & \multicolumn{1}{c}{$<$0.1} & \multicolumn{1}{c}{$<$0.1} \\
$\ttbar$                                    & 1.3,0.2  & 3.4,0.5 & 1.2,0.2 \\
Di-bosons + single top quark                           & 5.7,1.0  & 1.1,0.2 & 0.5, 0.1 \\
SM Higgs boson                                          & 3.7,1.3  & 0.6,0.2 & 0.2,0.1 \\
\hline
Total expected                              &    163.9,11.4 & 28.6,3.2 & 5.2,1.1\\
\hline
Observed data                                    & \multicolumn{1}{c}{165} & \multicolumn{1}{c}{26} & \multicolumn{1}{c}{1} \\
\end{scotch}

\vspace*{4ex}

\begin{scotch}{lxxx}
\multicolumn{4}{c}{Resonant analysis (event yields)} \\
\hline
Process & \multicolumn{1}{c}{0 \PQb tags} & \multicolumn{1}{c}{1 \PQb tag} & \multicolumn{1}{c}{2 \PQb tags} \\
\hline
500\GeV radion $\to \PH\PH$                                  & 1.6,0.2 & 5.7,0.7 & 6.2,0.8 \\
500\GeV graviton $\to \PH\PH$                              & 2.4,0.3 & 7.8,0.9 & 7.6,0.9\\
\hline
$\cPZ \to \Pgt\Pgt$                           &  130.6,13.8 & 19.8,3.4 & 2.7,1.0\\
Multijet                                 &   92.7,8.1 & 12.6,2.2 & 1.8,0.6\\
$\PW$+jets                                  &    8.4,1.5 & 0.8,0.3 & 0.4,0.1\\
$\cPZ$+jets ($\Pe$, $\Pgm$ or jet misidentified as $\tauh$) & 1.6,0.5 & \multicolumn{1}{c}{$<$0.1} & 0.2,0.1 \\
$\ttbar$                                   &2.5,0.4 & 5.2,0.7 & 2.7,0.5\\
Di-bosons + single top                        &  6.1,1.1 & 1.7,0.4 & 0.5,0.1\\
SM Higgs boson                                    &    5.0,1.7 & 0.7,0.2 & 0.2,0.1\\
\hline
Total expected                                &  246.8,13.9 & 40.6,3.9 & 8.4,1.3\\
\hline
Observed data                                    & \multicolumn{1}{c}{268} & \multicolumn{1}{c}{39} & \multicolumn{1}{c}{4} \\
\end{scotch}
\label{tab:eventYield}
\end{table*}

\begin{figure*}[htbp]
\centering
\includegraphics[width=0.48\textwidth]{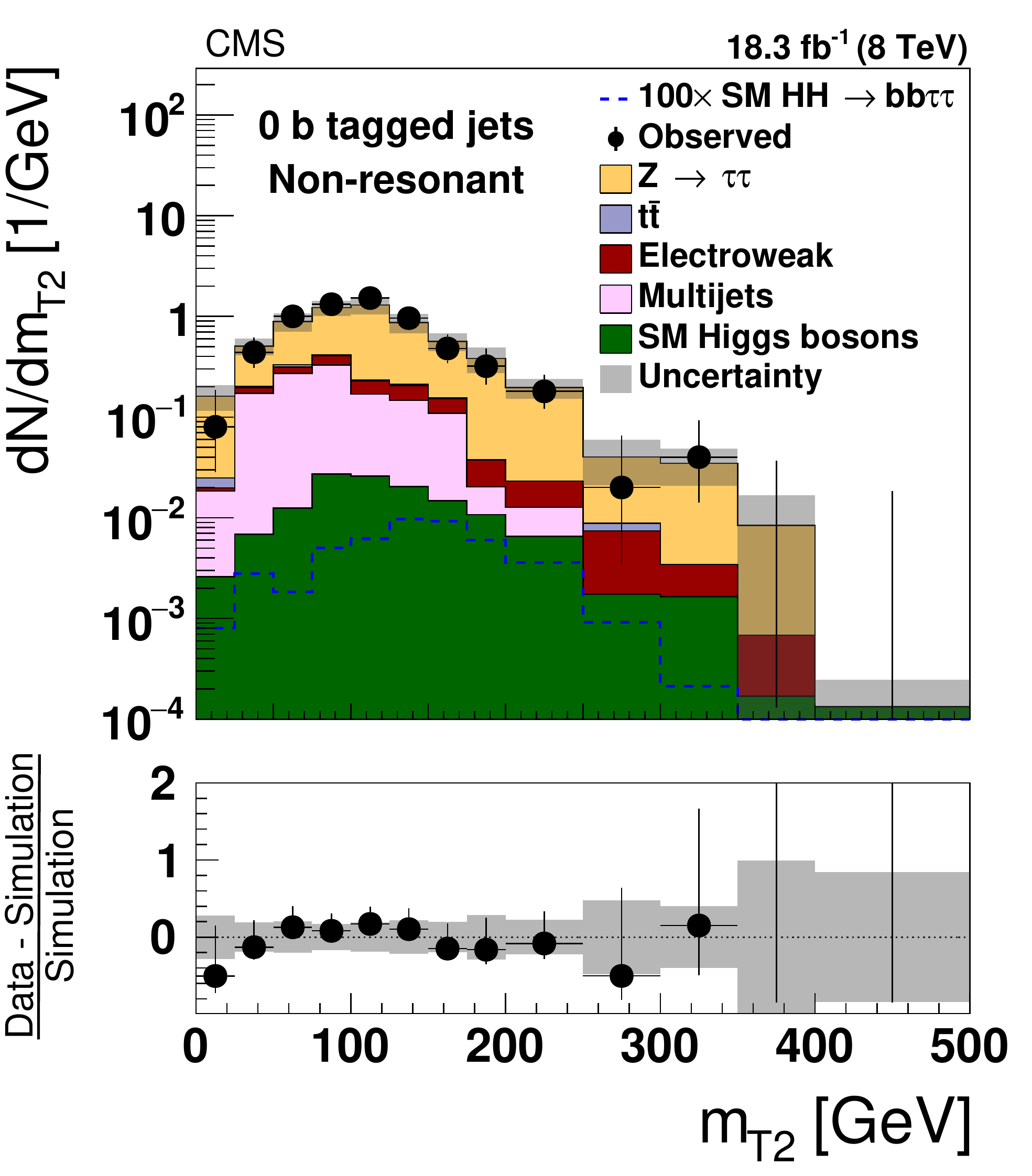}
\includegraphics[width=0.48\textwidth]{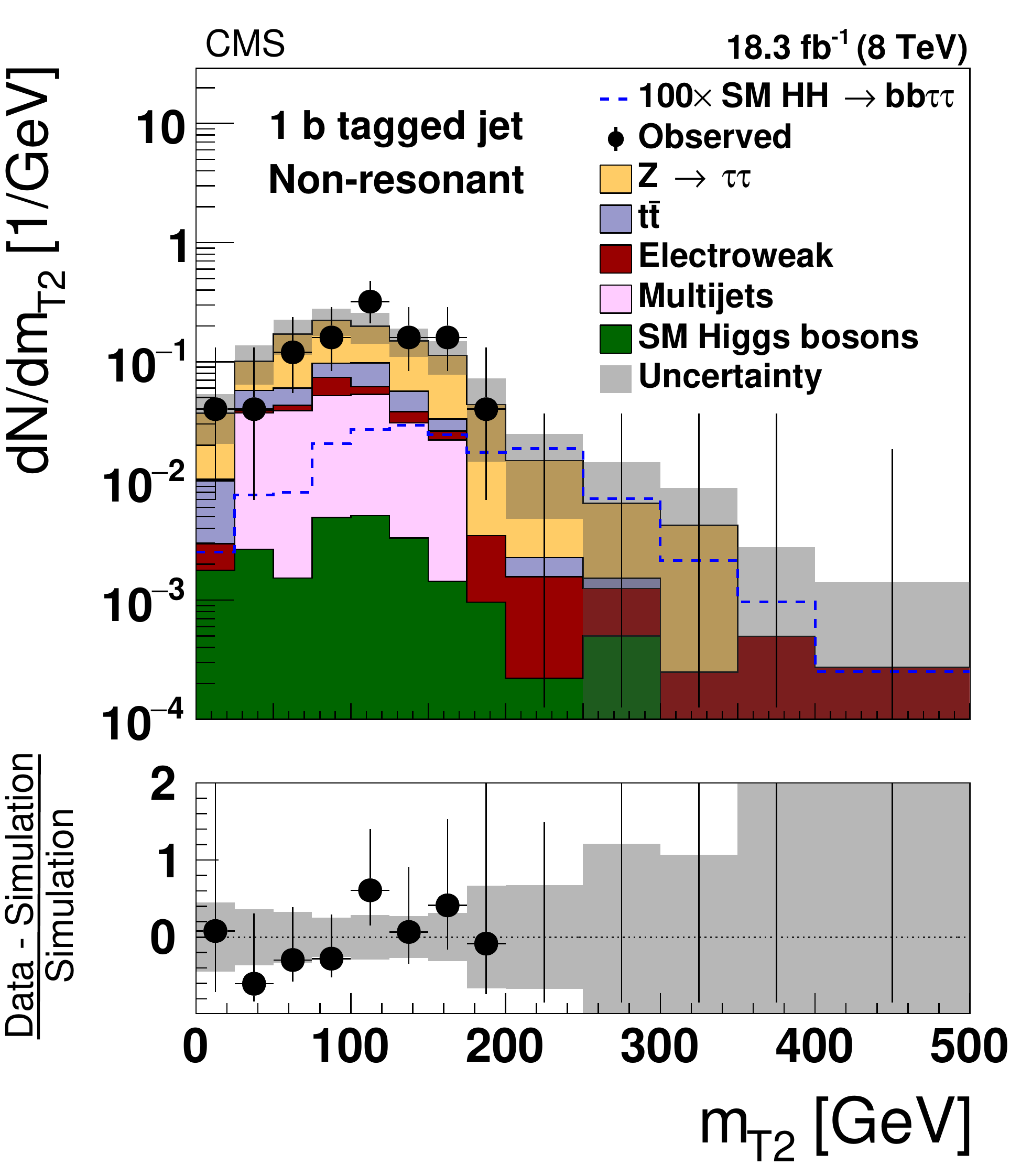}
\includegraphics[width=0.48\textwidth]{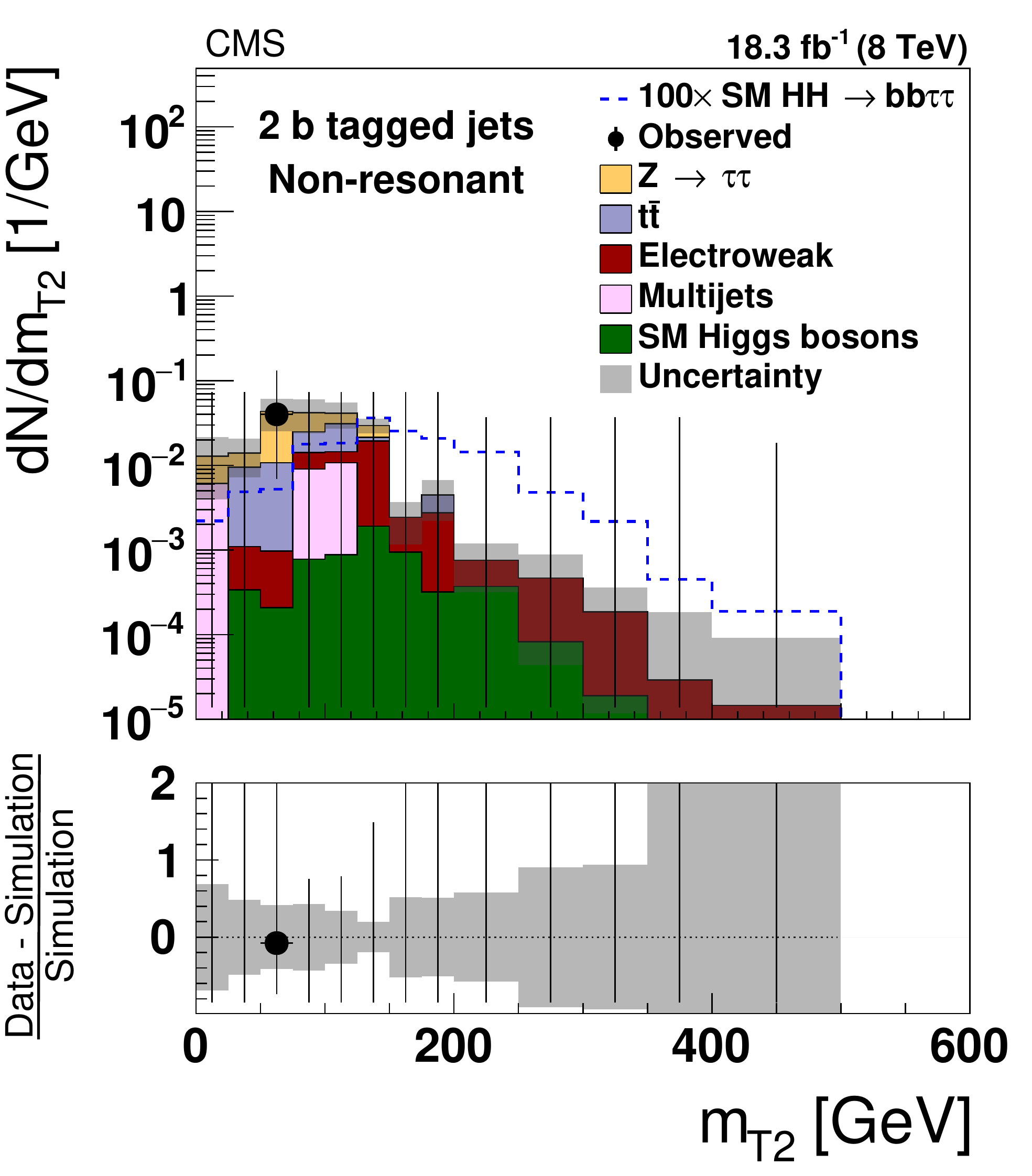}
\caption{
  Distributions in $m_{\mathrm{T2}}$ observed in the event categories with 0 \PQb tags, 1 \PQb tag, and 2 \PQb tags in the data
compared to the background expectation.
Hypothetical non-resonant $\PH\PH$ signals with a cross section
$\sigma(\Pp\Pp \to \PH\PH)$ of 1\unit{pb},
  corresponding to 100 times the SM cross section are overlaid for
comparison. The expectation for
signal and background processes is shown for values of nuisance parameters
 obtained from the likelihood fit.}
\label{fig:resultsMassDistributions2}
\end{figure*}

\begin{figure*}[htbp]
\centering
\includegraphics[width=0.48\textwidth]{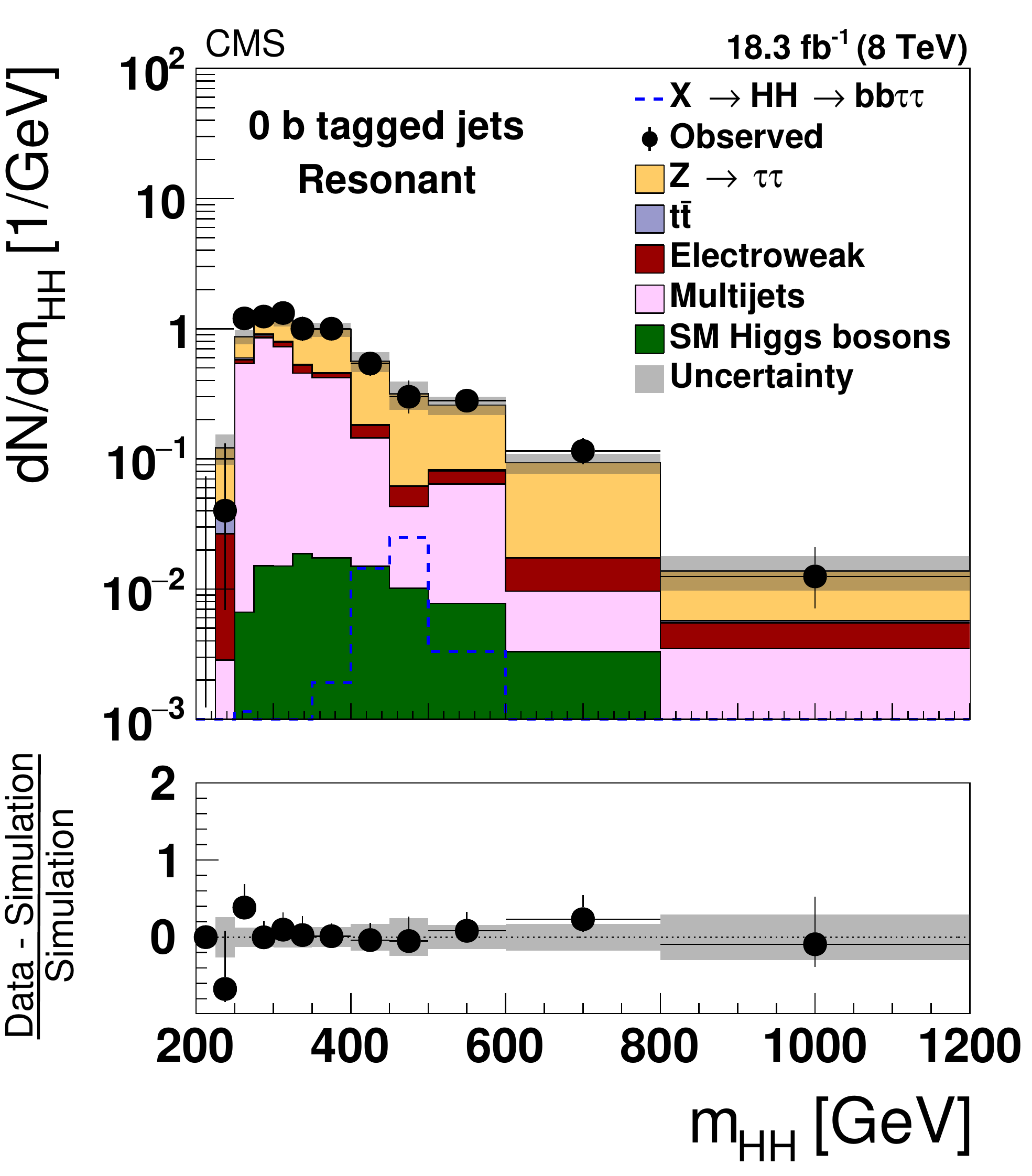}
\includegraphics[width=0.48\textwidth]{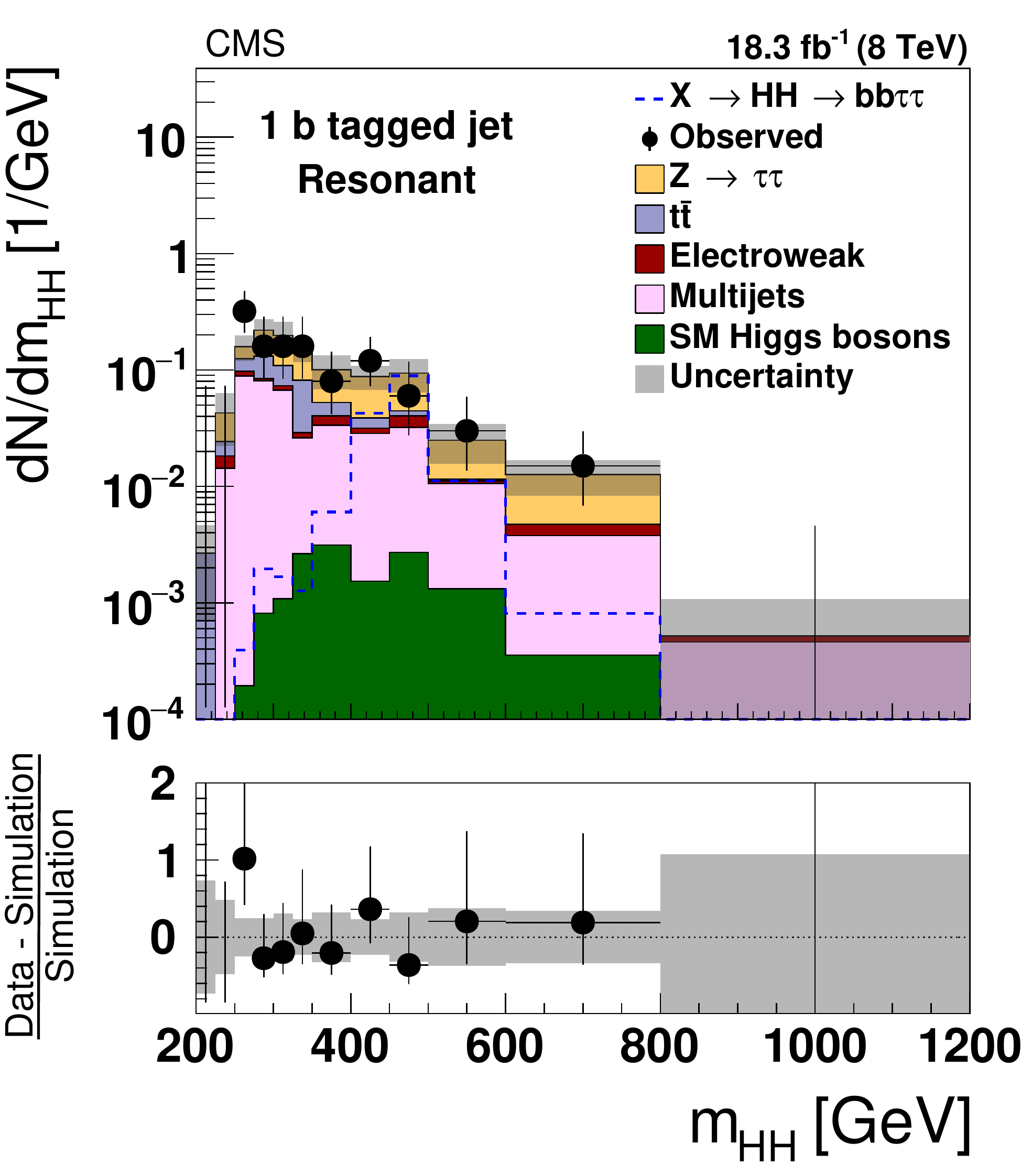}
\includegraphics[width=0.48\textwidth]{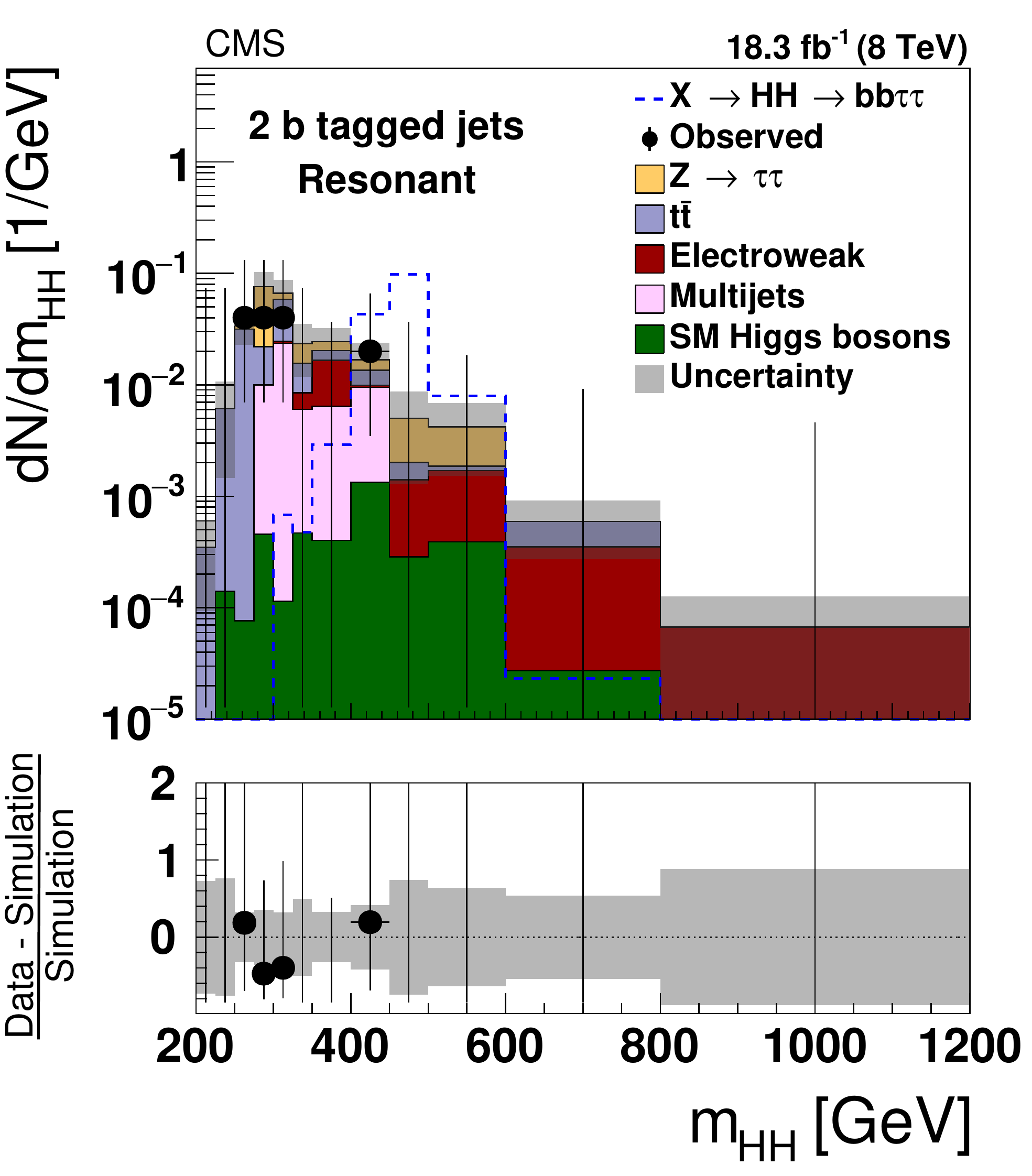}
\caption{
 Distributions in $m_{\PH\PH}$ observed in the event categories with 0 \PQb tags, 1 \PQb tag, and 2 \PQb tags in the data
compared to the background expectation.
Hypothetical signal distributions corresponding to the
 decays of a spin 2 resonance $\cPX$ of mass $m_{\cPX} = 500$\GeV
  that is produced with a
$\sigma(\Pp\Pp \to \cPX) \, \mathcal{B}(\cPX \to
\PH\PH))$ of 1\unit{pb} are overlaid for comparison.
The corresponding WED model parameters are $kl$ = 35 and
$k/{{\overline{M}}_{\mathrm{Pl}}}$ = 0.2.
 The expectation for signal and background processes is shown for values of
 nuisance parameters obtained from the likelihood fit.
}
\label{fig:resultsMassDistributions1}
\end{figure*}

For non-resonant $\PH\PH$ production the distributions of $m_{\textrm{T2}}$
are shown in Fig.~\ref{fig:resultsMassDistributions2}. For the resonant case
 the distribution of $m_{\PH\PH}$ for
 events selected in the three categories mentioned above are shown in
Fig.~\ref{fig:resultsMassDistributions1}.
In both figures, the sum of
$\PW$+jets, single top quark and di-boson events and of $\cPZ$+jets events 
in which one or both
 reconstructed $\tauh$ are due to a misidentified
$\Pe$, $\Pgm$, or jet is referred to as ``electroweak'' background.
Bins in which zero events are observed in the data are indicated by the absence of a
data point. The vertical bar drawn in these bins indicate the
84\% confidence interval, corresponding to a tail probability of 16\%.
The event yields and the shape of mass distributions
observed in data are in agreement with background predictions. No evidence
for the presence of a signal is observed.

\subsection{Cross section limits}
\label{sec:results_xSection}

\begin{figure}
\centering
\includegraphics[width=0.48\textwidth]{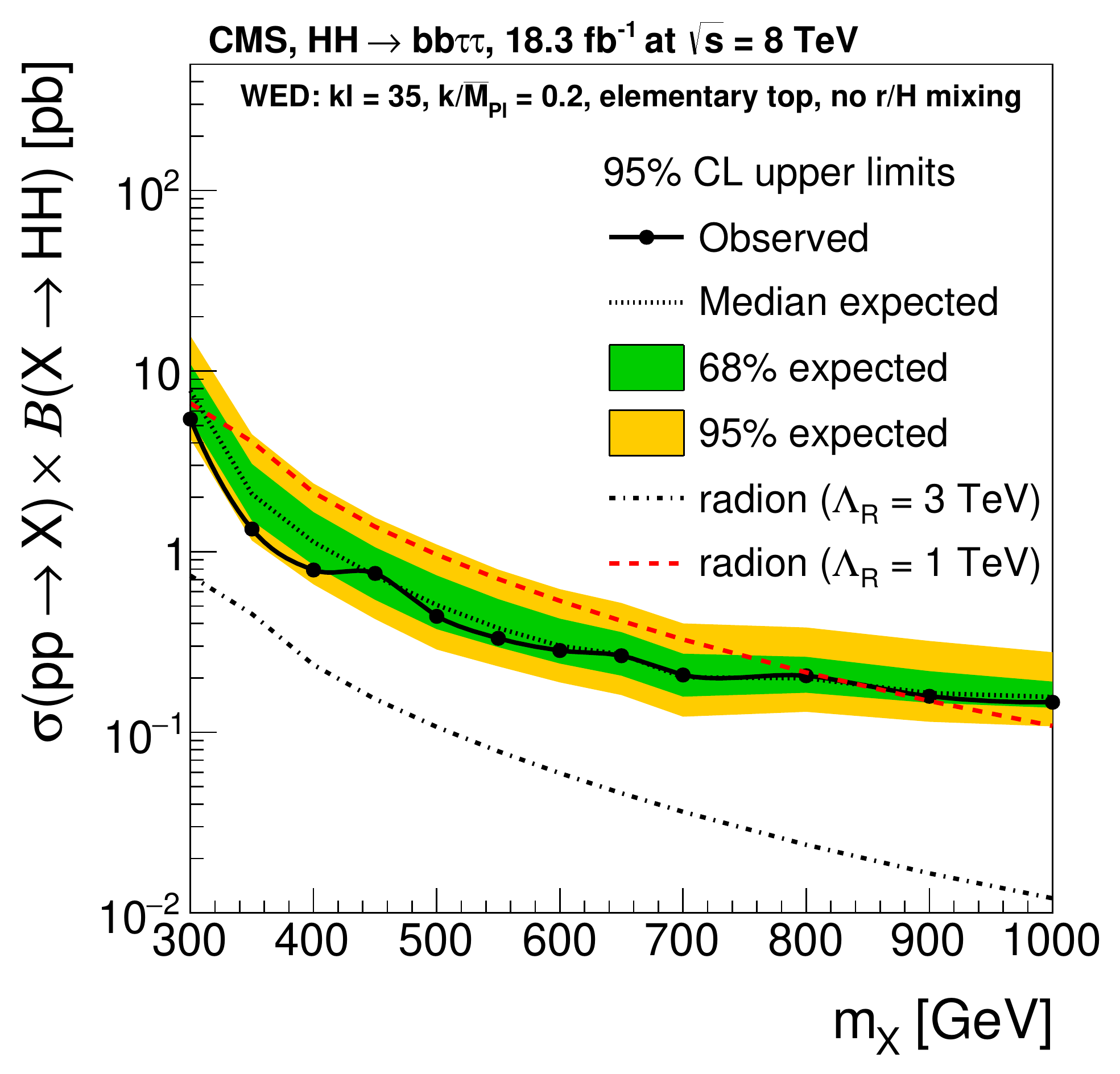}
\includegraphics[width=0.48\textwidth]{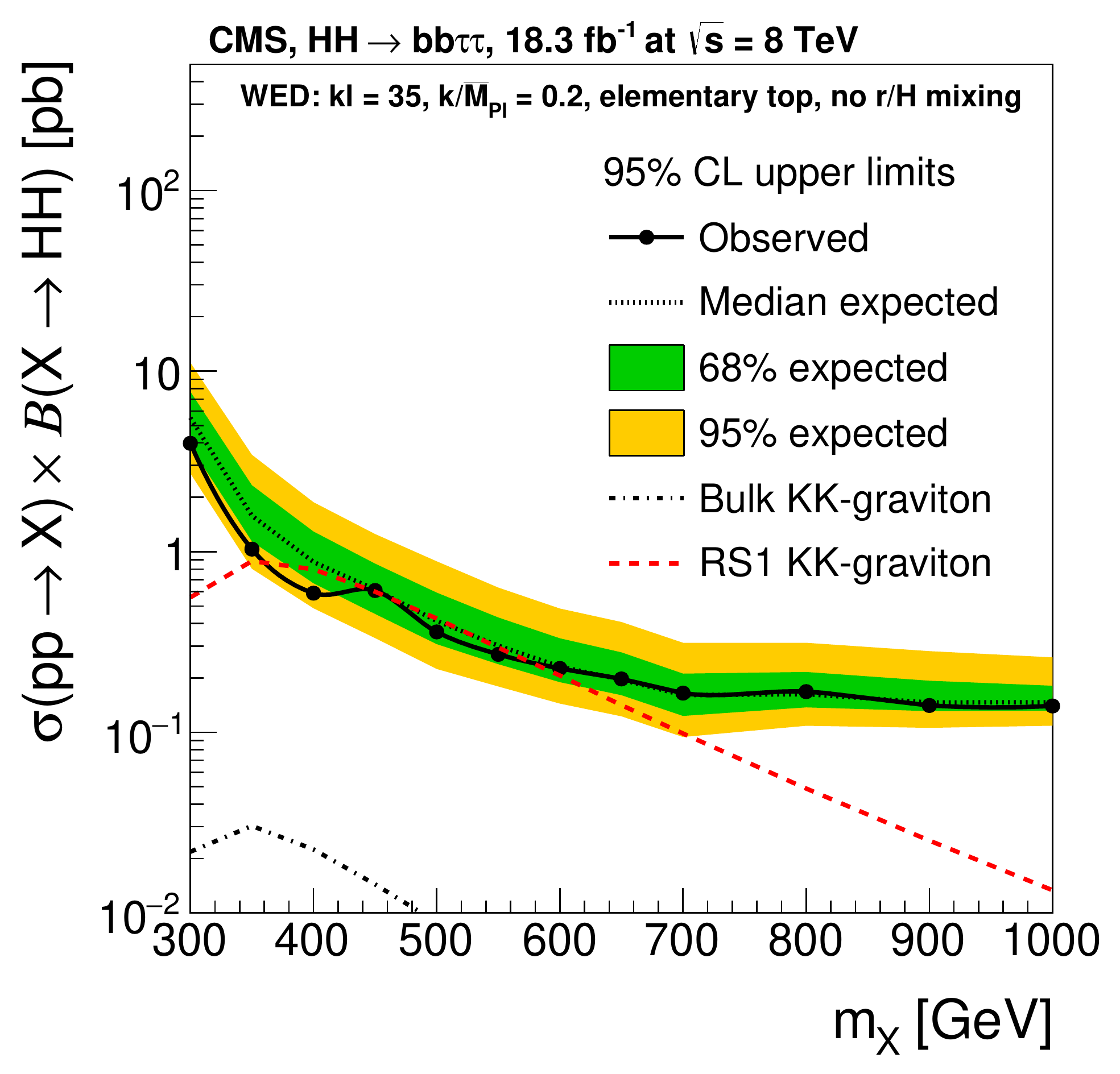}
\caption{
The 95\% \CL observed and expected upper limits on the
$\sigma(\Pp\Pp \to \cPX) \,
\mathcal{B}(\cPX \to \PH\PH$)
 for a spin 0 (\cmsLeft) and for a spin 2 (\cmsRight) resonance $\cPX$ as functions of the resonance mass $m_{\cPX}$, obtained from the search in the decay channel $\PQb\PQb\tau\tau$. The green and yellow
bands represent, respectively, the 1 and 2 standard deviation extensions beyond the expected limit. Also shown are theoretical predictions corresponding
to WED models for radions for values of
${\Lambda}_{\mathrm{R}} = 1, 3$\TeV and for RS1 and bulk KK gravitons~\cite{agashe, aquino}. The other WED model parameters are $kl = 35$ and
$k/{{\overline{M}}_{\mathrm{Pl}}} = 0.2$, assuming an elementary top
 hypothesis and no radion--Higgs (r/\PH) mixing.
}
\label{fig:results_limits_resonant}
\end{figure}

We have set 95\% \CL upper limits on cross section times
 branching fraction for $\PH\PH$ production using a modified frequentist approach,
known as the CL$_{\mathrm{s}}$
method~\cite{Junk, Read:2002hq, ATL-PHYS-PUB-2011-011}. For non-resonant
production SM event kinematics have been assumed.
Some model dependency is expected in this case, as the signal
 acceptance times efficiency as well as the shape of the $m_{\mathrm{T2}}$
 distribution vary as functions of the $m_{\PH\PH}$ spectrum
predicted by the model.
The observed (expected) limits on $\sigma(\Pp\Pp \to \PH\PH)$ are
0.59 ({0.94} $_{-0.24}^{+0.46}$)\unit{pb},
corresponding to a factor of about 59\,(94) times the cross section predicted
 by the SM. For the production of resonances decaying to a pair of SM-like
\PH's
 of mass $m_{\PH} = 125$\GeV the difference between the limits computed
for radion $\to \PH\PH$ and graviton $\to \PH\PH$ signals is small,
indicating that the limits on resonant $\PH\PH$ production cross section do not depend on these particular models. The limits obtained for resonant $\PH\PH$ production are given in Table~\ref{tab:results_limits} and are shown in Fig.~\ref{fig:results_limits_resonant}.
In this figure, the expected limits are computed for a generic spin 0/2 resonance decaying to two SM \PH's.
 The theoretical curves for the graviton case are based on KK graviton production in the bulk and RS1
models, respectively~\cite{agashe, aquino}. To obtain the radion theoretical
 curves, cross section for radion
production via gluon fusion are computed (to NLO electroweak and NNLO QCD
accuracy) for different
values of the fundamental theoretical parameter $\Lambda_{\mathrm{R}}$. These values are then multiplied by a k factor
calculated for SM-like $\PH$ production through gluon-gluon fusion
~\cite{giudice, mahanta, hdavoudiasl}.

\begin{table*}[htbp]
\topcaption{
  The 95\% \CL upper limits on resonant $\PH\PH$ production
($\sigma(\Pp\Pp \to \cPX) \,
\mathcal{B}(\cPX \to \PH\PH)$) in units of pb for spin 0 (radion)
 and spin 2 (graviton) resonances X, at different masses $m_{\cPX}$,
  obtained from the $\PH\PH$ search in the decay channel $\PQb\PQb\tau\tau$.
}
\centering
\begin{scotch}{l c c c c}
\multirow{2}{*}{$m_{\cPX}$ [\GeVns{}]} & \multicolumn{2}{c}{Radion (spin 0) ($\sigma$)} & \multicolumn{2}{c}{Graviton (spin 2) ($\sigma$)} \\
 & Expected (pb)& Observed (pb)& Expected (pb)& Observed (pb)\\
\hline
300  & 7.78 & 5.42 & 5.51 & 3.97 \\
350  & 2.08 & 1.33 & 1.58 & 1.03 \\
400  & 1.13 & 0.79 & 0.87 & 0.58 \\
450  & 0.73 & 0.75 & 0.61 & 0.60 \\
500  & 0.50 & 0.44 & 0.41 & 0.36 \\
600  & 0.30 & 0.28 & 0.23 & 0.23 \\
700  & 0.20 & 0.21 & 0.16 & 0.16 \\
800  & 0.19 & 0.20 & 0.16 & 0.16 \\
900  & 0.16 & 0.16 & 0.14 & 0.14 \\
1000 & 0.15 & 0.14 & 0.14 & 0.14 \\
\end{scotch}
\label{tab:results_limits}
\end{table*}

The results of the search for $\PH\PH$ production in the $\PQb\PQb\tau\tau$ decay channel
are combined with those in the decay channels $\gamma\gamma{\PQb}{\PQb}$ and $\PQb\PQb\PQb\PQb$,
published in Refs.~\cite{CMSsearch_2a, CMSsearch_2b} respectively.
The combination is performed by adding the three individual log likelihood
functions. The correlated systematics are taken into account by using the
same nuisance parameters for the fully correlated sources. They are
the luminosity uncertainty, the uncertainty on the \PQb tagging efficiency,
 the uncertainties related to the underlying event and parton showering, the
uncertainties on the branching fractions of the three $\PH\PH$ decays channels, and
the theoretical uncertainties on the SM non-resonant $\PH\PH$ cross section, parton density functions and ${\alpha}_{S}$.
The uncertainty on the branching fraction of $\PH\to{\gamma\gamma}$ is $\pm$5\%~\cite{Heinemeyer:2013tqa}.

The signal yield in the three decay channels is determined assuming that the branching fractions for the decays $\PH \to \PQb\PQb$, $\PH \to \tau\tau$, and $\PH \to \gamma\gamma$ are
equal to the SM predictions~\cite{Heinemeyer:2013tqa} for a $\PH$ with mass $m_{\PH} = 125$\GeV.
The datasets analyzed by the $\gamma\gamma{\PQb}{\PQb}$ and $\PQb\PQb\PQb\PQb$ decay channels
correspond to integrated luminosities of 19.7 and 17.9\fbinv,
recorded at $\sqrt{s} = 8$\TeV respectively.
The search in the $\gamma\gamma{\PQb}{\PQb}$ decay channel targets resonant as well as non-resonant $\PH\PH$ production,
while the search in the $\PQb\PQb\PQb\PQb$ decay channel focuses on resonant $\PH\PH$ signals. No evidence for a signal is observed in the combined search.

\begin{figure}
\centering
\includegraphics[width=0.48\textwidth]{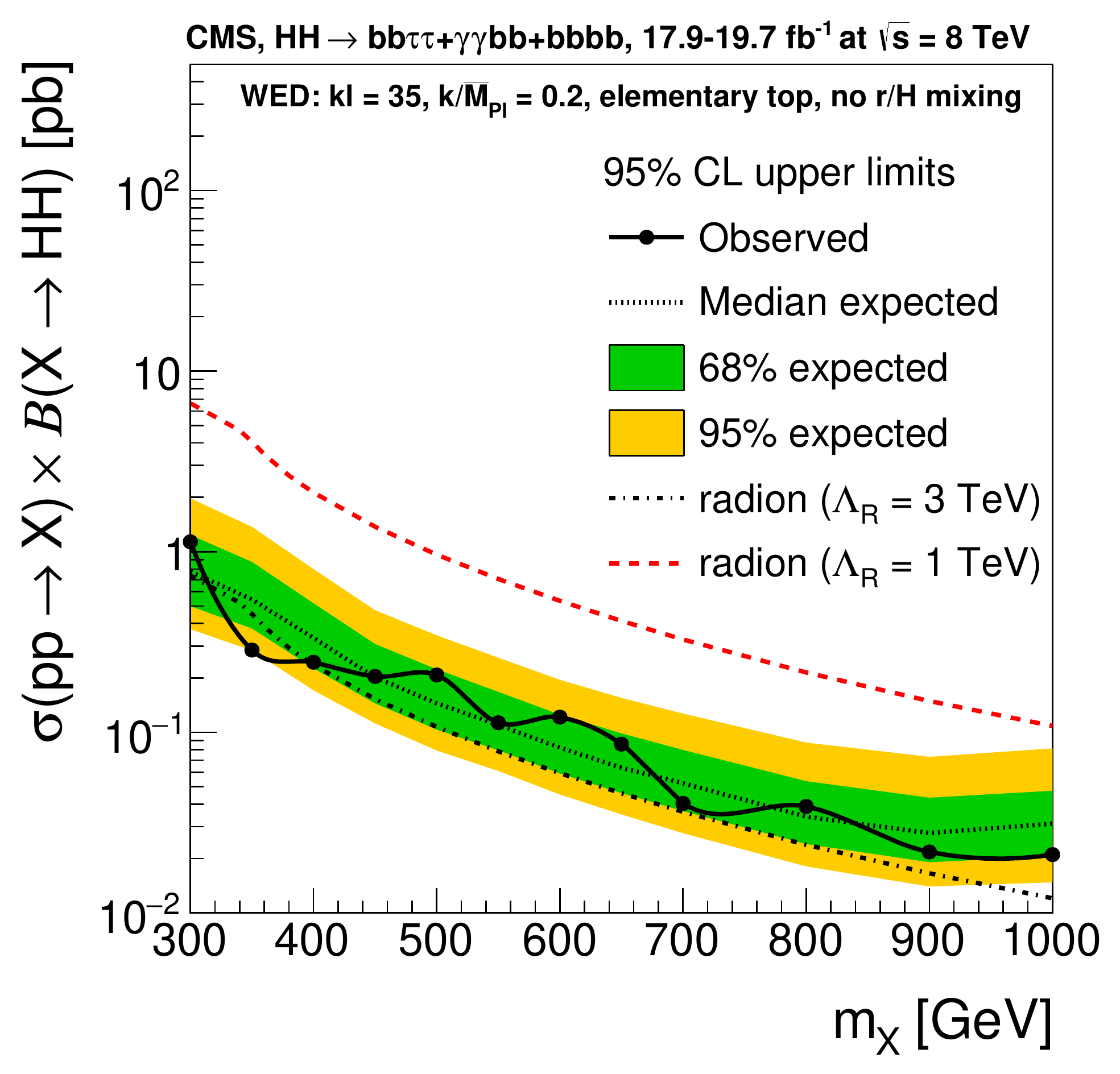}
\includegraphics[width=0.48\textwidth]{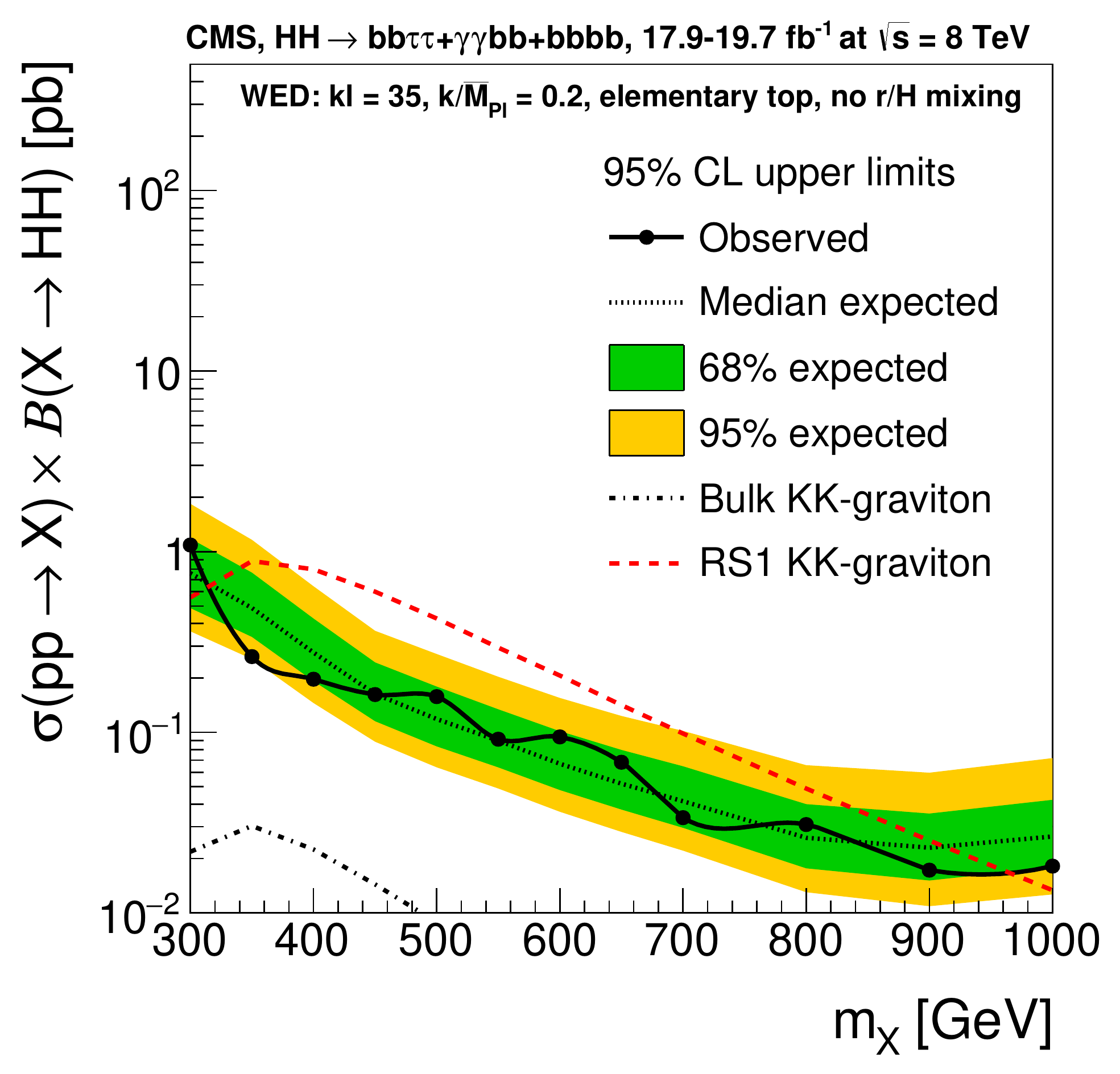}
\caption{95\% \CL observed and expected upper limits on the cross section times
branching fraction ($\sigma(\Pp\Pp \to \cPX) \, \mathcal{B}(\cPX \to \PH\PH)$)
 for a spin 0 (\cmsLeft) and for a spin 2 (\cmsRight) resonance $\cPX$ as functions of the resonance mass $m_{\cPX}$, obtained from the combination of searches performed in the $\PQb\PQb\tau\tau$, $\gamma\gamma{\PQb}{\PQb}$ and $\PQb\PQb\PQb\PQb$ decay channels. The green and yellow bands represent, respectively, the 1 and 2 standard deviation extensions beyond the expected limit. Also shown are theoretical predictions corresponding to WED models for radions for values of
${\Lambda}_{\mathrm{R}} = 1$, 3 and for RS1 and Bulk
 KK gravitons~\cite{agashe, aquino}. The other WED model parameters are
$kl = 35$ and $k/{{\overline{M}}_{\mathrm{Pl}}} = 0.2$, assuming an elementary
top hypothesis and no radion--Higgs (r/\PH) mixing.
}
\label{fig:results_limits_resonant_combination}
\end{figure}

The limits on resonant $\PH\PH$ production obtained from the combination of $\PQb\PQb\tau\tau$,
$\gamma\gamma{\PQb}{\PQb}$, and $\PQb\PQb\PQb\PQb$ channels are given in Table~\ref{tab:results_limits_combination} and Fig.~\ref{fig:results_limits_resonant_combination}.
In the case of non-resonant $\PH\PH$ production, an
 observed (expected) limit on $\sigma(\Pp\Pp \to \PH\PH)$ of
0.43\unit{pb} ({0.47} $_{-0.12}^{+0.20}$\unit{pb}),
corresponding to 43\,(47) times the SM cross section, is obtained by
 combining the $\PQb\PQb\tau\tau$ and
$\gamma\gamma{\PQb}{\PQb}$ decay channels. The low mass sensitivity
 ($m_{\PH\PH} \leq400$\GeV) is dominated by the
$\gamma\gamma{\PQb}{\PQb}$ channel while the
high mass ($m_{\PH\PH}> 700$\GeV) sensitivity is driven by the
$\PQb\PQb\PQb\PQb$ channel. The
 $\PQb\PQb\tau\tau$ channel is competitive with the
$\gamma\gamma{\PQb}{\PQb}$ channel
 in the intermediate mass range ($400\GeV< m_{\PH\PH} \leq 700$\GeV).

\begin{table*}[htbp]
\topcaption{
  The 95\% \CL upper limits on resonant $\PH\PH$ production
($\sigma(\Pp\Pp \to \cPX) \, \mathcal{B}(\cPX \to
\PH\PH)$) in units of fb for spin 0 (radion) and spin 2 (graviton)
resonances X, at different masses $m_{\cPX}$,
  obtained from the combination of $\PH\PH$ searches performed in the
$\PQb\PQb\tau\tau$, $\gamma\gamma{\PQb}{\PQb}$, and
$\PQb\PQb\PQb\PQb$ decay channels.
}
\centering
\begin{scotch}{ l c c c c }
\multirow{2}{*}{$m_{\cPX}$ [\GeVns{}]} & \multicolumn{2}{c}{Radion (spin 0) ($\sigma$)} & \multicolumn{2}{c}{Graviton (spin 2) ($\sigma$)} \\
 & Expected (fb)& Observed (fb)& Expected (fb)& Observed (fb)\\
\hline
300  & 776 & 1134 & 760 & 1088 \\
350  & 544 &  285 & 488 &  262 \\
400  & 333 &  244 & 276 &  197 \\
450  & 201 &  204 & 163 &  162 \\
500  & 145 &  207 & 118 &  157 \\
600  &  82 &  121 &  67 &  94 \\
700  &  52 &   40 &  41 &  34 \\
800  &  34 &   39 &  26 &  31 \\
900  &  28 &   22 &  23 &  17 \\
1000 &  31 &   21 &  26 &  18 \\
\end{scotch}
\label{tab:results_limits_combination}
\end{table*}

\section{Summary}
\label{sec:summary}

A search has been performed for events containing a pair of SM-like \PH's in resonant and non-resonant production of the pair in the channel where one boson decays to a pair of $\PQb$ quarks and the other to
a $\tau$ lepton pair, in $\Pp\Pp$ collisions collected by the CMS experiment at 8\TeV center-of-mass energy,
corresponding to an integrated luminosity of 18.3\fbinv.
Results are expressed as 95\% \CL upper limits on the production of a signal.
The limit on non-resonant $\PH\PH$ production corresponds to a factor of 59
 times the rate expected in the SM.
For resonant $\cPX$ $\to$ $\PH\PH$ production, the limit on
$\sigma(\Pp\Pp \to X) \, {\mathcal{B}}(\cPX \to \PH\PH)$ for a
resonance of spin 0 and spin 2 ranges, respectively, from
5.42 and 3.97\unit{pb} at a mass $m_{\cPX}=300$\GeV to 0.14\unit{pb} and
0.14\unit{pb} at $m_{\cPX}=1000$\GeV.

The results of the search in the $\PQb\PQb\tau\tau$ decay channel are
 combined with those in the $\gamma\gamma{\PQb}{\PQb}$
and $\PQb\PQb\PQb\PQb$ decay channels. For
non-resonant $\PH\PH$ production, the combination of $\PQb\PQb\tau\tau$
and $\gamma\gamma{\PQb}{\PQb}$ decay channels yields a limit that
 is a factor of 43 times the SM rate. The limit on resonant $\PH\PH$ production obtained from the
combination ranges from 1.13 and 1.09\unit{pb} at
 $m_{\cPX}=300\GeV$, to 21 and 18\unit{fb} at
$m_{\cPX}=1000\GeV$ for resonances of spin 0 and spin 2
respectively.

\begin{acknowledgments}
We congratulate our colleagues in the CERN accelerator departments for the excellent performance of the LHC and thank the technical and administrative staffs at CERN and at other CMS institutes for their contributions to the success of the CMS effort. In addition, we gratefully acknowledge the computing centers and personnel of the Worldwide LHC Computing Grid for delivering so effectively the computing infrastructure essential to our analyses. Finally, we acknowledge the enduring support for the construction and operation of the LHC and the CMS detector provided by the following funding agencies: BMWFW and FWF (Austria); FNRS and FWO (Belgium); CNPq, CAPES, FAPERJ, and FAPESP (Brazil); MES (Bulgaria); CERN; CAS, MoST, and NSFC (China); COLCIENCIAS (Colombia); MSES and CSF (Croatia); RPF (Cyprus); SENESCYT (Ecuador); MoER, ERC IUT, and ERDF (Estonia); Academy of Finland, MEC, and HIP (Finland); CEA and CNRS/IN2P3 (France); BMBF, DFG, and HGF (Germany); GSRT (Greece); OTKA and NIH (Hungary); DAE and DST (India); IPM (Iran); SFI (Ireland); INFN (Italy); MSIP and NRF (Republic of Korea); LAS (Lithuania); MOE and UM (Malaysia); BUAP, CINVESTAV, CONACYT, LNS, SEP, and UASLP-FAI (Mexico); MBIE (New Zealand); PAEC (Pakistan); MSHE and NSC (Poland); FCT (Portugal); JINR (Dubna); MON, RosAtom, RAS, RFBR and RAEP (Russia); MESTD (Serbia); SEIDI, CPAN, PCTI and FEDER (Spain); Swiss Funding Agencies (Switzerland); MST (Taipei); ThEPCenter, IPST, STAR, and NSTDA (Thailand); TUBITAK and TAEK (Turkey); NASU and SFFR (Ukraine); STFC (United Kingdom); DOE and NSF (USA).

{\tolerance=800
\hyphenation{Rachada-pisek} Individuals have received support from the Marie-Curie program and the European Research Council and Horizon 2020 Grant, contract No. 675440 (European Union); the Leventis Foundation; the A. P. Sloan Foundation; the Alexander von Humboldt Foundation; the Belgian Federal Science Policy Office; the Fonds pour la Formation \`a la Recherche dans l'Industrie et dans l'Agriculture (FRIA-Belgium); the Agentschap voor Innovatie door Wetenschap en Technologie (IWT-Belgium); the Ministry of Education, Youth and Sports (MEYS) of the Czech Republic; the Council of Science and Industrial Research, India; the HOMING PLUS program of the Foundation for Polish Science, cofinanced from European Union, Regional Development Fund, the Mobility Plus program of the Ministry of Science and Higher Education, the National Science Center (Poland), contracts Harmonia 2014/14/M/ST2/00428, Opus 2014/13/B/ST2/02543, 2014/15/B/ST2/03998, and 2015/19/B/ST2/02861, Sonata-bis 2012/07/E/ST2/01406; the National Priorities Research Program by Qatar National Research Fund; the Programa Clar\'in-COFUND del Principado de Asturias; the Thalis and Aristeia programs cofinanced by EU-ESF and the Greek NSRF; the Rachadapisek Sompot Fund for Postdoctoral Fellowship, Chulalongkorn University and the Chulalongkorn Academic into Its 2nd Century Project Advancement Project (Thailand); and the Welch Foundation, contract C-1845.
\par}
\end{acknowledgments}
\clearpage
\bibliography{auto_generated}
\cleardoublepage \appendix\section{The CMS Collaboration \label{app:collab}}\begin{sloppypar}\hyphenpenalty=5000\widowpenalty=500\clubpenalty=5000\textbf{Yerevan Physics Institute,  Yerevan,  Armenia}\\*[0pt]
A.M.~Sirunyan, A.~Tumasyan
\vskip\cmsinstskip
\textbf{Institut f\"{u}r Hochenergiephysik,  Wien,  Austria}\\*[0pt]
W.~Adam, E.~Asilar, T.~Bergauer, J.~Brandstetter, E.~Brondolin, M.~Dragicevic, J.~Er\"{o}, M.~Flechl, M.~Friedl, R.~Fr\"{u}hwirth\cmsAuthorMark{1}, V.M.~Ghete, C.~Hartl, N.~H\"{o}rmann, J.~Hrubec, M.~Jeitler\cmsAuthorMark{1}, A.~K\"{o}nig, I.~Kr\"{a}tschmer, D.~Liko, T.~Matsushita, I.~Mikulec, D.~Rabady, N.~Rad, B.~Rahbaran, H.~Rohringer, J.~Schieck\cmsAuthorMark{1}, J.~Strauss, W.~Waltenberger, C.-E.~Wulz\cmsAuthorMark{1}
% ...
\vskip\cmsinstskip
% ...
1:~~Also at Vienna University of Technology, Vienna, Austria\\
% ...

\end{sloppypar}
\end{document}